\documentclass{article}
\usepackage{amssymb}
\usepackage{amsmath}
\usepackage{pstricks}
\allowdisplaybreaks
\psset{unit=0.8cm}
\numberwithin{equation}{section}
\DeclareMathOperator{\pa}{\phantom a}
\DeclareMathOperator{\diag}{\rm diag}
\DeclareMathOperator*{\res}{\rm res}
\DeclareMathOperator{\Rs}{\mathbb{R}}
\DeclareMathOperator{\Cs}{\mathbb{C}}
\DeclareMathOperator{\Zs}{\mathbb{Z}}

\DeclareMathOperator{\Lo}{\mathcal{L}}
\DeclareMathOperator{\No}{\mathcal{N}}
\DeclareMathOperator{\Do}{\mathcal{D}}
\DeclareMathOperator{\Vo}{\mathcal{V}}

\DeclareMathOperator{\bk}{\mathbf{k}}
\def\on#1#2{\genfrac{}{}{0pt}{1}{#1}{#2}}
\newtheorem{theorem}{Theorem}[section]
\newtheorem{lemma}[theorem]{Lemma}

\newtheorem{definition}{Definition}[section]

\newtheorem{corollary}{Corollary}[section]
\begin{document}

\title{Heat operator with pure soliton potential: properties of Jost and dual Jost solutions}
\author{M.~Boiti${}^{*}$, F.~Pempinelli${}^{*}$, and A.~K.~Pogrebkov$
{}^{\dag}$ \\
${}^{*}$Dipartimento di Fisica, Universit\`a del Salento and\\
Sezione INFN, Lecce, Italy\\
${}^{\dag}$Steklov Mathematical Institute, Moscow, Russia}
\date{PACS: 02.30Ik, 02.30Jr, 05.45Yv}
\maketitle

\begin{abstract}
Properties of Jost and dual Jost solutions of the heat equation, $\Phi(x,\bk)$ and $\Psi (x,\bk)$, in the case of a pure solitonic
potential are studied in detail. We describe their analytical properties on the spectral parameter $\bk$ and their asymptotic behavior on the $x$-plane and we show that the values of $e^{-qx}\Phi(x,\bk)$ and the residua of $e^{qx}\Psi(x,\bk)$ at special discrete values of $\bk$ are bounded functions of $x$ in a polygonal region of the $q$-plane. Correspondingly, we deduce that the extended version $L(q)$ of the heat operator with a pure solitonic potential has left and right annihilators for $q$ belonging to these polygonal regions.
\end{abstract}

\section{Introduction}

The Kadomtsev--Petviashvili equation in its version called KPII
\begin{equation}
(u_{t}-6uu_{x_{1}}+u_{x_{1}x_{1}x_{1}})_{x_{1}}=-3u_{x_{2}x_{2}},\label{KPII}
\end{equation}
where $u=u(x,t)$, $x=(x_{1},x_{2})$ and subscripts $x_{1}$, $x_{2}$ and $t$ denote partial derivatives, is a (2+1)-dimensional generalization of the celebrated Korteweg--de~Vries (KdV) equation. There are two inequivalent versions of the KP equations, corresponding to the two choices $\pm$ for the sign in the rhs of~(\ref{KPII}), which are referred to, respectively, as KPI and
KPII. The KP equations, originally derived as a model for small-amplitude, long-wavelength, weakly two-dimensional waves in a weakly dispersive medium~\cite{KP1970}, were known to be integrable since the beginning of the 1970s~\cite{D1974,ZS1974}, and can be considered as prototypical (2+1)-dimensional integrable equations.

The KPII equation is integrable via its association to the operator
\begin{equation}
\Lo(x,\partial_{x})=-\partial_{x_{2}}+\partial_{x_{1}}^{2}-u(x),\label{heatop}
\end{equation}
which defines the well known equation of heat conduction, or heat equation for short, since it can be expressed as compatibility condition $[\Lo,\mathcal{T}]=0$ of the Lax pair $\Lo$ and $\mathcal{T}$ , where $\mathcal{T}$ is given by
\begin{equation}
\mathcal{T}(x,\partial_{x},\partial_{t})=\partial_{t}+4\partial_{x_{1}}^{3}-6u\partial_{x_{1}}-3u_{x_{1}}-
3\partial_{x_{1}}^{-1}u_{x_{2}}.  \label{Lax}
\end{equation}

The spectral theory of the operator (\ref{heatop}) was developed in \cite{BarYacoov,Lipovsky,Wickerhauser,Grinevich0} in the case of a real potential $u(x)$ rapidly decaying at spatial infinity, which, however, is not the most interesting case, since the KPII equation was just proposed in \cite{KP1970} in order to deal with two dimensional weak transverse perturbation of the one soliton solution of the KdV. In fact, if $u_{1}^{}(t,x_{1}^{})$ obeys KdV, then $u(t,x_{1}^{},x_{2}^{})=u_{1}^{}(t,x_{1}^{}+\mu x_{2}^{}-3\mu_{}^{2}t)$
solves KPII for an arbitrary constant $\mu\in\Rs$. In particular, KPII admits a one soliton solution of the form
\begin{equation}
u(x,t)=-\dfrac{(\kappa_{1}-\kappa_{2})^{2}}{2}\text{\textrm{sech}}^{2}\Biggl[\dfrac{\kappa_{1}^{}- \kappa_{2}^{}}{2}x_{1}+\dfrac{\kappa_{1}^{2}-\kappa_{2}^{2}}{2}x_{2}-2(\kappa_{1}^{3}-\kappa_{2}^{3})t\Biggr],
\label{1-sol}
\end{equation}
where $\kappa_{1}$ and $\kappa_{2}$ are real, arbitrary constants.

A spectral theory of the heat operator (\ref{heatop}) that also includes solitons has to be built. In the case of a potential
$u(x)\equiv u_{0}(x)$ rapidly decaying at spatial infinity, according to~\cite{BarYacoov,Lipovsky,Wickerhauser,Grinevich0}, the main tools in building the spectral theory of the operator (\ref{1.2}), as in the one dimensional case, are the integral equations, whose solutions define the related Jost solutions. However, if the potential $u(x)$ does not decay at spatial infinity, as it is the case when line soliton solutions are considered, the integral equations of the decaying case are ill-defined and one needs a more general approach. In solving the analogous problem for the nonstationary Schr\"{o}dinger operator, associated to the KPI equation, the extended resolvent
approach was introduced~\cite{BPP2006b}. Accordingly, a spectral theory of the KPII equation that also includes solitons has to be investigated using the resolvent approach. In this framework it was possible to develop the inverse scattering transform for a solution describing one soliton on a generic background~\cite{BPPP2002}, and to study the existence of the (extended) resolvent for (some) multisoliton solutions~\cite{BPPP2009}.

However, the general case of $N$-solitons is still open. Following~\cite{BPP2006b}, the first step in building the inverse scattering for this case lies in building a Green's function $G(x,x',\bk)$ of the heat operator (\ref{heatop}) corresponding to the pure $N$-soliton potential $u(x) $ such that
\begin{equation}
e_{}^{i\bk(x_{1}'-x_{1}^{})+\bk^{2}(x_{2}'-x_{2}^{})}G(x,x',\bk)  \label{green}
\end{equation}
is bounded for any $x$ and any $\bk\in \Cs$. We expect this Green's function to be bilinear in the Jost solutions $\Phi(x,\bk)$ and $\Psi (x,\bk)$ of the heat operator and its dual and, therefore, we need, first, to study the general properties of these solutions. Once performed this study and obtained the Green's function $G$, the Jost solution $\widetilde{\Phi}$ for a potential given as a sum of two terms,
\begin{equation}
\widetilde{u}(x)=u(x)+u'(x),
\end{equation}
where $u(x)$ is a pure $N$-soliton potential and $u'(x)$ an arbitrary bidimensional decaying smooth perturbation, can be derived as
solution of the integral equation
\begin{equation}
\widetilde{\Phi}=\Phi+Gu'\widetilde{\Phi},
\end{equation}
that is well defined since, thanks to $u'$, it has a good behaving kernel $Gu'$. Analogously, for the Jost solution $\widetilde{\Psi}$
of the dual heat operator. Then, having at hands these building blocks, $\widetilde{\Phi}$ and $\widetilde{\Psi}$, one can assemble the inverse scattering theory.

The paper in organized as follows. In Sec.\ 2 we consider the multisoliton potential showing at large spaces $N_{b}$ ``ingoing'' rays and $N_{a}$ ``outgoing'' rays obtained in its general form in \cite{BPPPr2001a}, and we reformulate it in terms of $\tau$-functions as in the
review paper \cite{ChK2} and in \cite{equivKPII}. In Sec.\ 3 we consider the Jost solutions of the heat operator and their dual, already obtained in \cite{BPPPr2001a}, but they also expressed in terms of $\tau$-functions as in \cite{equivKPII}. In view of getting the asymptotic behavior of the Jost solutions, in Sec.\ 4 the asymptotic behavior at large $x$ of the multisoliton potential, studied in details in \cite{equivKPII}, is reviewed showing that the round angle at the origin can be divided in $\No=N_{a}+N_{b}$ angular sectors such that on their bordering rays the potential has constant soliton like behavior, while along directions inside the sectors the potential has an exponential decaying behavior, which is explicitly given. In Sec.\ 5, by using the results obtained for the potential, also the asymptotic behavior at large $x$ of the Jost and dual Jost solutions and their values or, accordingly, residues at the discrete points of the spectrum are obtained. In Sec. 6 by using the results obtained in the previous section we show that the values of $e^{-qx}\Phi(x,\bk)$ and the residue of $e^{qx}\Psi (x,\bk)$ at special discrete values of $\bk$ are bounded functions of $x$ in a polygonal region of the $q$-plane included inside the parabola $q_{2}=q_{1}^{2}$. Correspondingly, we conclude that the extended version $L(q)$ of the heat operator with a pure solitonic potential has left and right annihilators for $q$ belonging to these polygonal regions. It follows that the extended resolvent $M(q)$, inverse of $L(q)$, does not exists for $q$ belonging to these polygons. However, the Green's function $G(x,x',\bk)$, needed, as indicated above, for building the inverse scattering theory in the case of KPII solutions with solitons, can be obtained by a reduction procedure involving $q$ values outside the parabola.

\section{Multisoliton potentials and their properties}

\subsection{Main notations}

Soliton potentials are labeled by the two numbers (topological charges), $N_{a}$ and $N_{b}$, that obey condition
\begin{equation}
N_{a},N_{b}\geq 1,  \label{nanb}
\end{equation}
Let
\begin{equation}
\No=N_{a}+N_{b},  \label{Nnanb}
\end{equation}
so that $\No\geq 2$. We introduce the $\No$ real parameters
\begin{equation}
\kappa_{1}<\kappa_{2}<\ldots <\kappa_{\No},  \label{kappas}
\end{equation}
and the functions
\begin{equation}
K_{n}^{}(x)=\kappa_{n}^{}x_{1}^{}+\kappa_{n}^{2}x_{2}^{},\quad n=1,\ldots ,\No.  \label{Kn}
\end{equation}
Let
\begin{equation}
e^{K(x)}=\diag\{e^{K_{n}(x)}\}_{n=1}^{\No},  \label{eK}
\end{equation}
be a diagonal $\No\times{\No}$-matrix, $\Do$ a $\No\times{N_{b}}$ constant matrix with at least two nonzero maximal minors, and let $\Vo$ be an ``incomplete Vandermonde matrix,'' i.e., the $N_{b}\times\No$-matrix
\begin{equation}
\Vo=\left(\begin{array}{lll}
1 & \ldots & 1 \\
\vdots &  & \vdots \\
\kappa_{1}^{N_{b}-1} & \ldots & \kappa_{\No}^{N_{b}-1}
\end{array}\right) .  \label{W}
\end{equation}
Then, the soliton potential is given by
\begin{equation}
u(x)=-2\partial_{x_{1}}^{2}\log \tau (x),  \label{ux}
\end{equation}
where the $\tau$-function can be expressed as
\begin{equation}
\tau (x)=\det\bigl(\Vo e^{K(x)}\Do\bigr).  \label{tau}
\end{equation}
See the review paper \cite{ChK2}, references therein, and \cite{equivKPII} where the same notations have been used.

There exists (see \cite{BC2}, \cite{BPPP2009}, and \cite{equivKPII}) a dual representation for the potential in terms of the $\tau$-function
\begin{equation}
\tau '(x)=\det \left( \Do^{\,\prime}e^{-K(x)}\gamma\Vo^{\,\prime}\right) ,  \label{tau'}
\end{equation}
where $\Do^{\,\prime}$ is a constant $N_{a}\times\No$-matrix that like the matrix $\Do$ has at least two nonzero maximal minors and that is orthogonal to the matrix $\Do$ in the sense that
\begin{equation}
\Do^{\,\prime}\Do=0,  \label{d12}
\end{equation}
being the zero in the r.h.s.\ a $N_{a}\times {N_{b}}$-matrix, and where $\Vo^{\,\prime}$ is the $\No\times{N_{b}}$-matrix
\begin{equation}
\Vo^{\,\prime}=\left(\begin{array}{lll}
1 & \ldots & \kappa_{1}^{N_{a}-1} \\
\vdots &  & \vdots \\
1 & \ldots & \kappa_{\No}^{N_{a}-1}
\end{array}\right) ,  \label{sol'}
\end{equation}
and $\gamma$ the constant, diagonal, real $\No\times\No$-matrix
\begin{equation}
\gamma =\diag\{\gamma_{n}\}_{n=1}^{\No},\qquad \gamma_{n}=
\prod_{\on{n'=1}{n'\neq {n}}}^{\No}(\kappa_{n}-\kappa_{n'})^{-1}.\label{gamma}
\end{equation}

In order to study the properties of the potential and the Jost solutions, it is convenient to use an explicit representation for the determinants. By using the Binet--Cauchy formula for the determinant of a product of matrices we get
\begin{equation}
\tau (x)=\sum_{1\leq n_{1}<n_{2}<\cdots <n_{N_{b}}\leq {\No}}f_{n_{1},\ldots
,n_{N_{b}}}\prod_{l=1}^{N_{b}}e^{K_{n_{l}}(x)},  \label{tauf}
\end{equation}
with
\begin{equation}
f_{n_{1},n_{2},\ldots ,n_{N_{b}}}=V(\kappa_{n_{1}}^{},\ldots ,\kappa_{n_{N_{b}}}^{})\Do(n_{1},\ldots ,n_{N_{b}}),\label{f}
\end{equation}
where we used notation
\begin{equation}
V(\kappa_{1},\ldots ,\kappa_{\No})=\prod_{1\leq m<n\leq \No}(\kappa_{n}-\kappa_{m}),  \label{V}
\end{equation}
for the Vandermonde determinant and notation
\begin{equation}
\Do(n_{1},\ldots ,n_{N_{b}})=\det \left(\begin{array}{lll}
\Do_{n_{1},1} & \dots & \Do_{n_{1},N_{b}} \\
\vdots &  & \vdots \\
\Do_{n_{N_{b}},1} & \dots & \Do_{n_{N_{b}},N_{b}}
\end{array}\right) ,  \label{Do}
\end{equation}
for the maximal minors of the matrix $\Do$. Notice that the coefficients $f_{n_{1},n_{2},\ldots ,n_{N_{b}}}$ are invariant under permutations of the indices.

In what follows it is convenient to consider indices $n,n_{1},\ldots,n_{N_{b}}$ of parameters $\kappa_{n}$, coefficients $f_{n_{1},n_{2},\ldots,n_{N_{b}}}$, etc, to be running mod$\,\No$ throughout $\Zs$, i.e., let
\begin{equation}
n\rightarrow n\,(\text{mod}\No).  \label{modN}
\end{equation}

From (\ref{tauf}) it follows directly that condition
\begin{equation}
f_{n_{1},\ldots ,n_{N_{b}}}\geq 0,\quad \text{for all }1\leq n_{1}<n_{2}<\cdots <n_{N_{b}}\leq {\No},  \label{reg}
\end{equation}
is sufficient (see \cite{K}) for the regularity of the potential $u(x)$, i.e., for the absence of zeros of $\tau(x)$ on the $x$-plane. Thanks to (\ref{kappas}), (\ref{V}) and (\ref{f}) this condition is equivalent to the condition that all maximal minors of the matrix $\Do$ are non negative. In \cite{equivKPII} it was mentioned that condition (\ref{reg}) is also necessary for the regularity of a  potential under evolution with respect to an arbitrary number of higher times of the KP hierarchy. In \cite{K} it is suggested to decompose soliton solutions of KPII into subclasses, associated to the Schubert cells on Grassmanian. It is proved there that condition (\ref{reg}) is necessary for the regularity of the all solutions associated to a cell. However, the problem of finding necessary conditions for the regularity of the many soliton solution under evolution with respect to KPII only is still open.

Under condition (\ref{modN}) and thanks to (\ref{kappas}) we have the following lemma (see \cite{asympKPII}) that we use below.
\begin{lemma}\label{lemma1} Let
\begin{equation}
g_{l,m,n}=(\kappa_{l}-\kappa_{m})(\kappa_{n}-\kappa_{n+N_{b}})(\kappa_{l}+\kappa_{m}-\kappa_{n}-\kappa_{n+N_{b}}).  \label{l11}
\end{equation}
Then
\begin{equation}
g_{l,m,n}\geq 0\quad \text{for any}\quad n\in \Zs,\quad l=n,\ldots,n+N_{b},\quad m=n+N_{b},\ldots ,\No+n,  \label{l12}
\end{equation}
and equality takes place only when $l$ and $m$ independently take values $n$ or $n+N_{b}$ by mod$\,\No$.
\end{lemma}

Matrices $\Do$ and $\Do^{\,\prime}$ obey rather interesting properties, see \cite{equivKPII,asympKPII}. In particular, products ${\Do}^{\dag}\Do$ and $\Do^{\,\prime}{\Do^{\,\prime}}^{\dag}$, where $\dag$ denotes Hermitian conjugation of matrices (in fact, transposition here), are positive and then invertible, so there exist matrices (see~\cite{G1990})
\begin{equation}
\bigl(\Do\bigr)^{(-1)}=({\Do}^{\dag}\Do)^{-1}{\Do}^{\dag},  \qquad
\bigl(\Do^{\,\prime}\bigr)^{(-1)}={\Do^{\,\prime}}^{\dag}(\Do^{\,\prime}{\Do^{\,\prime}}^{\dag})^{-1},  \label{d15}
\end{equation}
that are, respectively, the left inverse of the matrix $\Do$ and the right inverse of the matrix $\Do^{\,\prime}$, i.e.,
\begin{equation}
\bigl(\Do\bigr)^{(-1)}\Do=E_{N_{b}},\qquad \Do^{\,\prime}\bigl(\Do^{\,\prime}\bigr)^{(-1)}=E_{N_{a}},  \label{d17}
\end{equation}
where $E_{N_{a}}$ and $E_{N_{b}}$ are, respectively, the $N_{a}\times N_{a}$ and $N_{b}\times N_{b}$ identity matrices. Products of these matrices in the opposite order give the real self-adjoint $\No\times\No$-matrices
\begin{align}
& P=\Do\bigl(\Do\bigr)^{(-1)}=\Do({\Do}^{\dag}\Do)^{-1}\bigl(\Do\bigr)^{\dag},  \label{d19} \\
& P^{\,\prime}=\bigl(\Do^{\,\prime}\bigr)^{(-1)}\Do^{\,\prime}=\bigl(\Do^{\,\prime}\bigr)^{\dag}
(\Do^{\,\prime}{\Do^{\,\prime}}^{\dag})^{-1}\Do^{\,\prime},  \label{d18}
\end{align}
which are orthogonal projectors, i.e.,
\begin{equation}
P^{2}=P,\qquad (P^{\,\prime})^{2}=P^{\,\prime},\qquad PP^{\,\prime}=0=P^{\,\prime}P,  \label{d22}
\end{equation}
and complementary in the sense that
\begin{equation}
P+P^{\,\prime}=E_{\No}.  \label{d23}
\end{equation}

Matrices $\Do$ and $\Do'$ are not in one-to-one correspondence with potential $u(x)$. Indeed, by (\ref{ux}), (\ref{tau}), (\ref{tau'}) one gets that the potential is invariant under substitutions
\begin{equation}
\Do\to\Do{v},\qquad\Do^{\,\prime}\to{v'}\Do^{\,\prime},\label{d23:1}
\end{equation}
where $v$ and $v'$ are arbitrary constant, nonsingular $N_b\times{N_b}$- and  $N_a\times{N_a}$-matrices, correspondingly. Thus under condition (\ref{kappas}) ($N_a,N_b$)-soliton potential is parameterized by the point of the Grassmanian $\text{Gr}_{N_b,\No}$, if representation (\ref{tau}) is used, or by the point of the Grassmanian $\text{Gr}_{N_a,\No}$, if representation (\ref{tau'}) is used.

\subsection{Asymptotic behavior of the $\tau$-function and potential $u(x)$}

Here we shortly report (see \cite{asympKPII} for details) the main asymptotic properties of the function $\tau(x)$  and potential $u(x)$ at large space, which are necessary to know in order to derive the asymptotic properties of the Jost and dual Jost solutions and of their discrete values.

The asymptotic of $\tau (x)$ is determined by the interrelations between functions $K_{n}(x)$ for different $n$, that is, due to (\ref{Kn}), by the differences
\begin{equation}
K_{l}(x)-K_{m}(x)=(\kappa_{l}-\kappa_{m})(x_{1}+(\kappa_{l}+\kappa_{m})x_{2}),\qquad \text{for any\ }
l,m\in \mathbb{Z},  \label{linen}
\end{equation}
which are linear with respect to the space variables. Then, the asymptotic behavior must have sectorial structure on the $x$-plane and in order to describe this structure we divide the $x$-plane in sectors.

\begin{figure}[tbp]
\begin{center}
\begin{pspicture}[showgrid=false,linewidth=1.0pt](-4.5,-4.5)(4.5,4.5)
\psline[linestyle=dashed]{->}(0,-4.3)(0,4.3)
\psline[linestyle=dashed]{->}(-4.3,0)(4.3,0)
\psline(3,4)(0,0)(1,4)
\psline(4,-1)(0,0)(4,0.5)
\psline(-4,-1.25)(0,0)(-4,0.3)
\psline[linestyle=dotted]{<->}(0.75,3.1)(1.4,3)(2,2.7)
\psline[linestyle=dotted]{<->}(-2.6,-0.8)(-2.85,-0.35)(-2.8,0.2)
\psline[linestyle=dotted]{<->}(2.8,-0.7)(2.95,-0.2)(2.75,0.32)
\rput(-3.2,-0.4){\small$\sigma^{}_{1}$}
\rput(1.5,3.2){\small$\sigma^{}_{n}$}
\rput(3.35,-0.35){\small$\sigma^{}_{N_a}$}
\uput[0](4.4,-0.4){$x^{}_1$}
\uput[90](-0.3,4.1){$x^{}_2$}
\uput[0](-3.6,-1.4){\small$r^{}_{1}$}
\uput[0](3.5,-1.2){\small$r^{}_{N_a-1}$}
\uput[0](3.5,0.75){\small$r^{}_{N_a}$}
\uput[0](3.5,3.75){\small$r^{}_{n-1}$}
\uput[0](0.7,3.75){\small$r^{}_{n}$}
\uput[0](-3.6,.8){\small$r^{}_{\No}$}
\end{pspicture}
\end{center}
\caption{Sector $\protect\sigma_{n}$ corresponds to $N_{a}+2\leq n\leq\No-1$.}
\label{fig}
\end{figure}
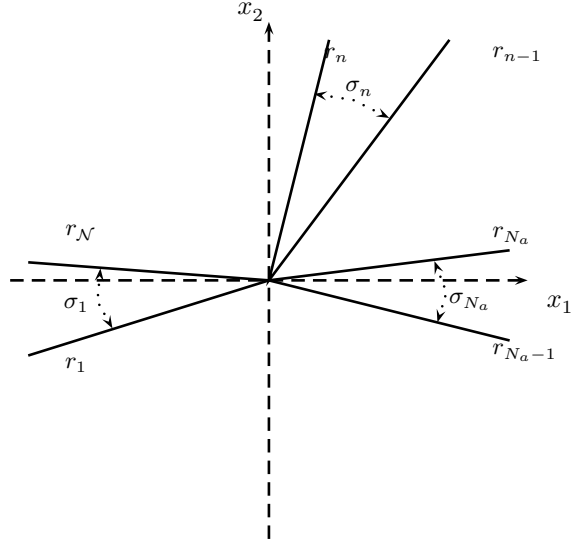

We introduce on the $x$-plane rays $r_{n}$ given by
\begin{equation}
r_{n}=\{x:x_{1}+(\kappa_{n+N_{b}}+\kappa_{n})x_{2}=0,\,(\kappa_{n+N_{b}}-\kappa_{n})x_{2}<0),\qquad
n=1,\ldots ,\No.  \label{rayn}
\end{equation}
They intersect the lines $x_{2}=\pm 1$, respectively, at the points
\begin{align}
&(\kappa_{n+N_b}+\kappa_n,-1) \qquad \text{for\ } n=1,\dots,N_a\notag\\
&(-\kappa_{n+N_b}-\kappa_n,1) \qquad \text{for\ } n=N_a+1,\dots,\No.\label{y0n}
\end{align}
Therefore, increasing $n$ from $n=1$ to $n=N_{a}$ the ray rotates anticlockwise in the lower half $x$-plane, crosses the positive part of the $x_{1}$-axis in coming to $n=N_{a}+1$ and, then, rotates anticlockwise in the upper half $x$-plane up to $n=\No$ and, finally, crossing the negative part of the $x_{1}$-axis, for $n=\No+1$ comes back to the ray $r_{1}(x)$ thanks to (\ref{modN}), see Fig.~\ref{fig}.

Let us assume that some rays, say, $r_{m}$ and $r_{n}$, where for definiteness $m<n$, are parallel. By (\ref{rayn}) this means that $\kappa_{m+N_{b}}+\kappa_{m}=\kappa_{n+N_{b}}+\kappa_{n}$, i.e., that $\kappa_{n}-\kappa_{m}=\kappa_{m+N_{b}}-\kappa_{n+N_{b}}$, where the l.h.s.\ is positive thanks to (\ref{kappas}). Then, because of (\ref{modN}) it is easy to see that the r.h.s.\ can be positive only if $1\leq {m}\leq {N_{a}}<{n}\leq \No$, so that by (\ref{y0n}) ray $r_{m}$ is in the bottom halfplane and $r_{n}$ in the upper one. We see that for generic values of the $\kappa_{n}$'s and when $N_{a}\neq N_{b}$ the rays in the upper and bottom half planes cannot be parallel, while this is possible for a special choice of $\kappa_{n}$'s. On the contrary, in the case $N_{a}=N_{b}$ all pairs $r_{n}$ and $r_{n+N_{a}}$, $n=1,\ldots ,N_{a}$, and only these pairs give parallel rays. In the special case $N_{a}=N_{b}=1$ we get two rays producing the straight line $x_{1}+(\kappa_{1}+\kappa_{2})x_{2}=0$ that divides the $x$-plane in two halfplanes.

In the same way we introduce sectors $\sigma_{n}$, which are subsets of the $x$-plane characterized as
\begin{equation}
\sigma_{n}=\{x:K_{n-1}(x)<K_{n+N_{b}-1}(x)\text{ and } K_{n}(x)>K_{n+N_{b}}(x)\},\quad \text{for } n=1,\dots ,\No.
\label{sigman}
\end{equation}
Since (\ref{linen}) and the discussion above it follows that the $\sigma_{n} $'s are sharp (for $\No>2$) angular sectors with vertices at the origin of the coordinates bounded from the right (looking from the origin) by the ray $r_{n-1}$ and from the left (looking from the origin) by the ray $r_{n}$ and that the sectors $\sigma_{n}$ with the increasing of $n$ are ordered anticlockwise, starting ``from the left'' with the sector $\sigma_{1}$, that includes the negative part of the $x_{1}$-axis, and, then, with the sectors $\sigma_{n}$ ($n=2,\ldots ,N_{a}$) in the bottom half-plane, the sector $\sigma_{N_{a}+1}$ ``to the right,'' that includes the positive part of the $x_{1}$-axis, and the sectors $\sigma_{n}$ ($n=N_{a}+2,\ldots ,\No$) on the upper half-plane, finishing with the sector $\sigma_{\No}$ tangent to the sector $\sigma_{1}$, covering in this way the whole $x$-plane with the exception of the bordering rays $r_{n}$, see Fig.~\ref{fig}.

Therefore, the sectors $\sigma_{n}$ define a $\No$-fold discretization of the round angle at the origin, with the integer $n$ (mod $\No$) playing the role of a discrete angular variable. It is clear that in the study of the asymptotic behavior of the $\tau $-function we consider the $x$-plane as a vector space, since the finite part of $x$ when $x\rightarrow \infty $ is irrelevant.

For determining the directions of rays $r_{n}$ and sectors $\sigma_{n}$, we introduce the vectors
\begin{equation}
y_{n}=(\kappa_{n+N_{b}}^{2}-\kappa_{n}^{2},\kappa_{n}-\kappa_{n+N_{b}}),\qquad n=1,\dots ,\No,  \label{pointn'}
\end{equation}
and use them for giving the following definition.

\begin{definition}
\label{def1'} We say that $x\rightarrow \infty $ along the ray $r_{n}$ if $x\rightarrow\infty $ and there exists such $\alpha\rightarrow+\infty$ that $x-\alpha y_{n}$ is bounded. This will be denoted as $x\overset{r_{n}}{\longrightarrow}\infty$.

We say that $x\rightarrow \infty $ in the sector $\sigma_{n}$, $n=1,\ldots,\No$, if $x\rightarrow\infty $ and there exist such $\alpha \rightarrow+\infty $ and $\beta \rightarrow +\infty$ that $x-\alpha y_{n-1}-\beta y_{n} $ is bounded. This will be denoted as $x\overset{\sigma_{n}}{\longrightarrow}\infty$.
\end{definition}

Notice that for $x\overset{r_{n}}{\longrightarrow}\infty$
\begin{equation}
K_{n}(x)-K_{n+N_{b}}(x)\quad \text{is bounded and}\qquad (\kappa_{n+N_{b}}-\kappa_{n})x_{2}\rightarrow -\infty ,  \label{def12}
\end{equation}
and for $x\overset{\sigma_{n}}{\longrightarrow}\infty$
\begin{equation}
K_{n+N_{b}-1}(x)-K_{n-1}(x)\rightarrow +\infty ,\qquad K_{n}(x)-K_{n+N_{b}}(x)\rightarrow +\infty ,  \label{def11}
\end{equation}
as follows directly from the definition. In fact we have a more general statement.

\begin{lemma}
\label{lemma3} Let $N_{a},N_{b}\geq 1$, and $n\in\Zs$ arbitrary. Then
\begin{enumerate}
\item if $x\overset{r_{n}}{\longrightarrow}\infty$ we have that $K_{l}(x)-K_{m}(x)\to+\infty$ or bounded for any $l=n,\ldots,n+N_{b}$ and $m=n+N_{b},\ldots,\No+n$, where boundedness takes place if and only if $(l,m)=(n,n)$, $(n,n+N_b)$, $(n+N_b,n)$, $(n+N_b,n+N_b)$;
\item if $x\overset{\sigma_{n}}{\longrightarrow}\infty$ we have that $K_{l}(x)-K_{m}(x)\to+\infty$ for any $l=n,\ldots,n+N_{b}-1$ and $m=n+N_{b},\ldots,\No+n-1$;
\end{enumerate}
where summation of indices is always understood mod $\No$.
\end{lemma}

Thanks to the above Lemma \ref{lemma3} the asymptotics of the $\tau $-function is given by the following Theorem (see \cite{asympKPII}).

\begin{theorem}
\label{th1} If condition (\ref{reg}) is satisfied, the asymptotic expansion of $\tau(x)$ for $x\rightarrow\infty$ for any $n\in\mathbb{Z}$ is given by
\begin{align}
&x\overset{r_{n}}{\longrightarrow}\infty: & &
\tau (x)=\bigl(z_{n}+z_{n+1}e_{}^{K_{N_{b}+n}(x)-K_{n}(x)}+o(1)\bigr)
\exp\left(\sum_{j=n}^{n+N_{b}-1}K_{j}(x) \right),  \label{3:232} \\
&x\overset{\sigma_{n}}{\longrightarrow}\infty: & &
\tau(x)=(z_{n}+o(1))\exp\left(\sum_{l=n}^{n+N_{b}-1}K_{l}(x)\right),\label{tauasympt}
\end{align}
where notation
\begin{equation}
z_{n}=f_{n,n+1,\ldots ,n+N_{b}-1}\equiv V(\kappa_{n}^{},\ldots ,\kappa_{n+N_{b}-1}^{})\Do(n,\ldots ,n+N_{b}-1),  \label{zn}
\end{equation}
was introduced for any $n=1,\ldots ,\No$.
\end{theorem}

As we see from (\ref{3:232}) and (\ref{tauasympt}) the leading asymptotic term along the ray direction $r_{n}$ is given by the sum of the leading terms obtained for $x\overset{\sigma_{n}}{\longrightarrow}\infty $ and
$x\overset{\sigma_{n+1}}{\longrightarrow}\infty $. Since the exponential factor cancels out when this expansion (\ref{3:232}) of $\tau(x)$ is inserted in~(\ref{ux}), the factor in parenthesis gives the ray behavior of the potential $u(x)$ at infinity. Explicitly, taking into account (\ref{modN}) we get $N_{a}$ asymptotic rays in the bottom half plane: $x_{2}\rightarrow-\infty $, $x_{1}+(\kappa_{n}+\kappa_{n+N_{b}})x_{2}$ bounded. Along these rays the potential behaves as
\begin{equation}
u(x)=-2\partial_{x_{1}}^{2}\log \bigl(z_{n}+z_{n+1}e_{}^{K_{n+N_{b}}(x)-K_{n}(x)}\bigr),\quad n=1,\ldots ,N_{a}.\label{down}
\end{equation}
In the same way, the potential $u(x)$ has $N_{b}$ asymptotic rays in the upper half-plane: $x_{2}\rightarrow +\infty $, $x_{1}+(\kappa_{n}+\kappa_{n+N_{a}})x_{2}$ bounded. Along these rays it behaves as
\begin{equation}
u(x)=-2\partial_{x_{1}}^{2}\log \bigl(z_{n+N_{a}}+z_{n+N_{a}+1}e_{}^{K_{n}(x)-K_{n+N_{a}}(x)}\bigr),\quad
n=1,\ldots ,N_{b}.  \label{up}
\end{equation}
Asymptotic behavior inside sectors $\sigma_{n}$ is given by (\ref{tauasympt}) where the only $x$-dependent term is the exponential factor. Taking into account its linear dependence on $x$ and (\ref{reg}) we get that $u(x)$ decays (exponentially) in all directions inside all sectors, i.e., on the whole $x$-plane with exception to the rays $r_1,\ldots,r_{\No}$.

Thanks to condition (\ref{modN})  coefficients $z_n$ defined in (\ref{zn}) are a special subset of coefficients
$f_{n_{1},\ldots ,n_{N_{b}}}$ in (\ref{f}). Theorem \ref{th1} shows that the behavior of the potential at large $x$  is determined by the coefficients $z_n$ only. Therefore, condition
\begin{equation}
z_{n}>0,\qquad n=1,\ldots ,\No,  \label{zn0}
\end{equation}
is a sufficient condition for the regularity of the potential at large $x$. If the stronger  condition (\ref{reg}) is not satisfied, singularities can appear, but only in a finite region of the $x$-plane.

Taking into account the special role played by coefficients $z_n$ let us introduce matrices
\begin{equation}
v_{n}=\left(\begin{array}{ccc}
\Do_{n,1}, & \ldots , & \Do_{n,N_{b}} \\
\ldots & \ldots & \ldots \\
\Do_{n+N_{b}-1,1}, & \ldots , & \Do_{n+N_{b}-1,N_{b}}
\end{array}\right) ,\qquad n=1,\dots,\No, \label{d61}
\end{equation}
so that by  (\ref{Do})  $\det {v_{n}}=\Do(n,\ldots ,n+N_{b}-1)$ and by (\ref{zn})
\begin{equation}
z_n=V(\kappa_{n,\ldots,\kappa_{n+N_b-1}})\det {v_{n}},\label{d61:1}
\end{equation}
that is different from zero thanks to (\ref{zn0}). Then, we can perform the special permutation
\begin{equation}
\pi_{n}(l)=N_{a}+1+l-n,\quad l=1,\ldots ,\No,  \label{d62}
\end{equation}
that shifts matrix $v_{n}$ in the bottom part of matrix $\Do$. Thus we get the representations
\begin{equation}
\Do=\pi_{n}\left(\begin{array}{c}
d_{(n)} \\
E_{N_{b}}
\end{array}\right) v_{n},\qquad
\Do^{\,\prime}=v_{n}'(E_{N_{a}},-d_{(n)})\pi_{n}^{\dag},  \label{blockzn}
\end{equation}
where the $N_{a}\times{N_{a}}$-matrix $v_{n}'$ is defined in analogy. Here $\pi_n$ denotes matrix of permutation (\ref{d62}), $\pi^{\dag}_{n}$ its Hermitially conjugate, $E_{N_a}$ and $E_{N_b}$ are, correspondingly, $N_{a}\times{N_{a}}$ and $N_{b}\times{N_{b}}$ unity matrices. The $N_a\times{N_b}$ rectangular submatrix $d_{(n)}$ is defined by any of the equalities in (\ref{blockzn}) and it is obvious that, thanks to their block structure, matrices $\Do$ and $\Do^{\,\prime}$ obey (\ref{d12}). Representation (\ref{blockzn}) is convenient for studying the asymptotic properties of the Jost solutions.

\section{Jost solutions}\label{Jost}

Since the heat operator is not self-dual, one has to consider simultaneously its dual
$\Lo^{d}(x,\partial_{x})=\partial_{x_{2}}+\partial_{x_{1}}^{2}-u(x)$ and, then, introduce the Jost solution $\Phi (x,\bk)$ and
the dual Jost solution $\Psi (x,\bk)$ obeying equations
\begin{equation}
\Lo(x,\partial_{x})\Phi (x,\bk)=0,\qquad \Lo^{d}(x,\partial_{x})\Psi (x,\bk)=0,  \label{1.2}
\end{equation}
where $\bk$ is an arbitrary complex variable, playing the role of a spectral parameter. The reality of the potential $u(x)$, that we always assume here, is equivalent to the conjugation properties
\begin{equation}
\overline{\Phi (x,\bk)}=\Phi (x,-\overline{\bk}),\qquad \overline{\Psi (x,\bk)}=\Psi (x,-\overline{\bk}).  \label{chixiconj}
\end{equation}
The $\tau$-function representations for the Jost solutions were derived in~\cite{equivKPII} and are given as
\begin{equation}
\Phi (x,\bk)=\dfrac{\tau_{\Phi}^{}(x,\bk)}{\tau (x)}e^{-i\bk x_{1}-\bk^{2}x_{2}},\qquad
\Psi (x,\bk)=\dfrac{\tau_{\Psi}^{}(x,\bk)}{\tau (x)}e^{i\bk x_{1}+\bk^{2}x_{2}},  \label{sym}
\end{equation}
with
\begin{equation}
\tau_{\Phi}(x,\bk)=\det \bigl(\Vo e^{K(x)}(\kappa +i\bk)\Do\bigr),\qquad
\tau_{\Psi}(x,\bk)=\det \bigl(\Vo e^{K(x)}(\kappa +i\bk)^{-1}\Do\bigr),\label{tauk}
\end{equation}
where $\kappa +i\bk$ denotes the diagonal $\No\times\No$-matrix
\begin{equation}
\kappa +i\bk=\diag\{\kappa_{1}^{}+i\bk,\ldots ,\kappa_{\No}^{}+i\bk\}\label{diagk}
\end{equation}
and analogously for the matrix $(\kappa +i\bk)^{-1}$. By this definitions $e^{i\bk x_{1}+\bk^{2}x_{2}}\Phi (x,\bk)$ is a polynomial with respect to $\bk$ of order $N_{b}$ and $e^{-i\bk x_{1}-\bk^{2}x_{2}}\Psi (x,\bk)$ is a meromorphic function of $\bk$ that becomes a polynomial of order $N_{a}$ after multiplication by $\prod_{n=1}^{\No}(\kappa_{n}+i\bk)$. In other words, $\Phi (x,\bk)$ is an entire function of $\bk$ and $\Psi(x,\bk)$ a meromorphic function with poles at points $\bk=i\kappa_{n}$, $n=1,\ldots ,\No$. Introducing the discrete values of $\Phi(x,\bk)$ at these points as a $\No$-row
\begin{equation}
\Phi (x,i\kappa)=\{\Phi (x,i\kappa_{1}^{}),\ldots ,\Phi (x,i\kappa_{\No}^{})\},  \label{Phik}
\end{equation}
and the residuals of $\Psi (x,\bk)$ at these points
\begin{equation}
\Psi_{\kappa_{n}}(x)=\res_{\bk=i\kappa_{n}}\Psi (x,\bk),  \label{resn}
\end{equation}
as a $\No$-column
\begin{equation}
\Psi_{\kappa}(x)=\{\Psi_{\kappa_{1}}(x),\ldots ,\Psi_{\kappa_{\No}}(x)\}^{\text{T}},  \label{Psik}
\end{equation}
we have the following relations (see \cite{equivKPII})
\begin{equation}
\Phi (x,i\kappa)\Do=0,\qquad \Do^{\,\prime}\Psi_{\kappa}(x)=0.\label{d6}
\end{equation}
Thanks to (\ref{d19}) and (\ref{d18}) relations (\ref{d6}) can be written equivalently in the form
$\Phi(x,i\kappa)P=0$, $P'\Psi_{\kappa}(x)=0$, so that by (\ref{d23}) and the last equality in (\ref{d22})
\begin{equation}
\Phi (x,i\kappa)=\Phi (x,i\kappa)(P+P')=\Phi (x,i\kappa)P',\qquad \Psi_{\kappa}(x)=(P+P')\Psi_{\kappa}(x)=P\Psi_{\kappa}(x), \label{d54}
\end{equation}
and, then, by~(\ref{d22}) for any $x$ and $x'$
\begin{equation}
\Phi (x,i\kappa)\Psi_{\kappa}(x')\equiv \sum_{n=1}^{\No}\Phi(x,i\kappa_{n})\Psi_{\kappa_{n}}(x')=0.  \label{hirota}
\end{equation}
This means that the product $\Phi (x,\bk)\Psi (x',\bk)$ of the Jost solutions obeys the well known Hirota bilinear identity~\cite{Miwa} for the Beiker--Akhiezer solutions, if the contour of integration surrounds all points $\bk=i\kappa_{n}$.

Let us introduce
\begin{equation}
\varphi (x)=\Phi (x,i\kappa){\Do^{\,\prime}}^{(-1)},\qquad \psi (x)=\Do^{(-1)}\Psi_{\kappa}(x), \label{d55}
\end{equation}
where notation (\ref{d15}) was used. By this definition, we have
\begin{align}
& \varphi (x)=\bigl(\varphi_{1}(x),\ldots ,\varphi_{N_{a}}(x)\bigr),\quad N_{a}\text{-row},  \label{d56} \\
& \psi (x)=\bigl(\psi_{1}(x),\ldots ,\psi_{N_{b}}(x)\bigr)^{\text{T}},\quad N_{b}\text{-column},  \label{d57}
\end{align}
and, by (\ref{d19}) and (\ref{d18}), we get representations
\begin{equation}
\Phi (x,i\kappa)=\varphi (x)\Do^{\,\prime},\qquad \Psi_{\kappa}=\Do\psi(x).  \label{d58}
\end{equation}

Thus we constructed $N_{a}$ solutions and $N_{b}$ dual solutions that parameterize discrete values of the Jost and dual Jost solutions (that have $\No$ components). It is clear that $\varphi (x)$ and $\psi (x)$ are not invariant with respect to the redefinition of matrices $\Do$ and $\Do'$ mentioned in (\ref{d23:1}), while, in contrast, combinations $\Do{v_{n}^{-1}}$ and ${v_{n}'}^{-1}\Do'$ are invariant with respect to transformation (\ref{d23:1}), as follows from (\ref{blockzn}).

It is necessary to mention that with the above definitions (\ref{sym}) the Jost solutions have at large $\bk$ the asymptotics
\begin{equation}
\lim_{\bk\rightarrow \infty}(i\bk)^{-N_{b}}e^{i\bk x_{1}+\bk^{2}x_{2}}\Phi(x,\bk)=1,\qquad
\lim_{\bk\rightarrow \infty}(i\bk)^{N_{b}}e^{-i\bk x_{1}-\bk^{2}x_{2}}\Psi (x,\bk)=1,  \label{asymptk2}
\end{equation}
and the potential is reconstructed as
\begin{align}
u(x)& =-2\lim_{\bk\rightarrow \infty}(i\bk)^{-N_{b}+1}\partial_{x_{1}}\bigl(e^{i\bk x_{1}+\bk^{2}x_{2}}\Phi (x,\bk)\bigr)\equiv  \notag \\
& \equiv 2\lim_{\bk\rightarrow \infty}(i\bk)^{N_{b}+1}\partial_{x_{1}} \bigl(e^{-i\bk x_{1}-\bk^{2}x_{2}}\Psi (x,\bk)\bigr).  \label{asymptk3}
\end{align}

\section{Asymptotics of the Jost solutions}

\subsection{Asymptotics of the Jost solutions $\Phi(x,\bk)$ and $\Psi(x,\bk)$ for a generic $\bk\in\Cs$}

Since, as was already noted in \cite{equivKPII}, the functions $\tau_{\Phi}(x,\bk)$ and $\tau_{\Psi}(x,\bk)$ defined in~(\ref{tauk}) can be obtained from the function $\tau (x)$ by means, respectively, of the special Miwa shifts~\cite{Miwa}
$e^{K(x)}\rightarrow {e}^{K(x)}(\kappa +i\bk)$ and $e^{K(x)}\rightarrow {e}^{K(x)}(\kappa +i\bk)^{-1}$, their asymptotic behavior follows trivially from Theorem~\ref{th1} for $\tau (x)$ if $\bk\neq i\kappa_{n}$ for all $n$. Therefore, from equation~(\ref{sym}) we get that the Jost solutions for $x\overset{\sigma_{n}}{\longrightarrow}\infty $ inside an arbitrary angular sector $\sigma_{n}$, $n=1,\ldots ,\No$, have the following asymptotic behaviors:
\begin{align}
\Phi (x,\bk)& =e^{-i\bk x_{1}-\bk^{2}x_{2}}\prod_{j=n}^{n+N_{b}-1}(\kappa_{j}+i\bk)+\ldots ,  \label{Phik1} \\
\Psi (x,\bk)& =e^{i\bk x_{1}+\bk^{2}x_{2}}\prod_{j=n}^{n+N_{b}-1}(\kappa_{j}+i\bk)^{-1}+\ldots ,  \label{Psik1}
\end{align}
were summation of indices is by mod $\No$ and were dots denote weaker terms. Asymptotic behavior along the ray, $x\overset{r_{n}}{\longrightarrow}\infty$ also easily follows from Theorem~\ref{th1} and we omit it here, since in what
follows we are interested only in the exponential behavior of the leading terms, while in the ray directions we have two leading terms with the same asymptotic behavior (see (\ref{3:232})). To emphasize this we introduce the ``closed'' sectors $\overline{\sigma^{\pa}_{n}}$. In other words we introduce:

\begin{definition}
\label{def3} We say that $x\to\infty$ along the direction of the closed sector $\overline{\sigma^{\pa}_{n}}$, $n=1,\ldots,\No$, if $x\to\infty$ and there exist non negative $\alpha$ and $\beta$ such that $x-\alpha y_{n-1}-\beta y_{n}$ is bounded when $\alpha+\beta\to\infty$.  This will be denoted as $x\overset{\overline{\sigma^{\pa}_{n}}}{\longrightarrow}\infty$.
\end{definition}

According to this definition, asymptotics (\ref{Phik1}) and (\ref{Psik1}) are valid for $x\overset{\overline{\sigma^{\pa}_{n}}}{\longrightarrow}\infty$.

In what follows it is enough to study the asymptotic behavior of only one of the Jost solutions, $\Phi(x,\bk)$ or $\Psi(x,\bk)$. Indeed, thanks to (\ref{sym}), the dual representations for $\tau $ in (\ref{tau}) and (\ref{tau'}) and the analogous ones for
$\tau_{\Phi}$ and $\tau_{\Psi}$ (see Remark 5.1 in \cite{equivKPII}), one can derive that the Jost solutions $\Phi (x,\bk)$ and $\Psi (x,\bk)$ are related by the equation
\begin{equation}
\Psi (x,\bk)=\bigl[\Phi (-x,\bk)\bigr]\prod_{n=1}^{\No}(\kappa_{n}+i\bk)^{-1},  \label{-PhiPsi}
\end{equation}
where square brackets in the r.h.s.\ means that one has to perform in $\Phi(-x,\bk)$, as defined in (\ref{tauk}), the substitutions
\begin{equation}\label{PP}
N_{a}\leftrightarrow N_{b},\qquad \Do\rightarrow \gamma \Do^{\,\prime \text{T}},
\end{equation}
where matrices $\Do^{\,\prime}$ and $\gamma $ are defined in (\ref{d12}) and (\ref{gamma}) and superscript T denotes matrix transposition. It is easy to see that asymptotics (\ref{Phik1}) and (\ref{Psik1}) obey this property, if one notice that, thanks to Definitions \ref{def1'}, substitution (\ref{PP}) leads to the following relations for the sectors:
\begin{equation}
[\sigma_{n}]=\sigma_{n+N_{a}},\qquad [\sigma_{n+N_{a}}]=\sigma_{n},  \label{sigmasigma}
\end{equation}
so that asymptotic behavior of $\Psi (x,\bk)$ in sector $\sigma_{n}$ is given by the asymptotic behavior of $\Phi(-x,\bk)$ in sector $\sigma_{n+N_{a}}$, once performed the transformations (\ref{PP}).

When $\bk=i\kappa_{m}$, for some $m$, some terms in $\tau_{\Phi}(x,\bk)$ (see (\ref{tauk})) are absent. In the case when such terms are coefficients of the leading exponents we get zero asymptotics and exact behavior of the Jost solution cannot be derived from (\ref{Phik1}). The same is valid for residuals of $\Psi(x,\bk)$. Thus we have to consider asymptotic behavior at these values of $\bk$ separately.

\subsection{Asymptotics of the discrete values of the Jost solutions}

In order to find the asymptotic behavior of the discrete values of the Jost solutions, we notice that according to~(\ref{sym}), (\ref{tauk}) and definition~(\ref{Kn}), they equal
\begin{equation}
\Phi (x,i\kappa_{m})=\dfrac{\tau_{\Phi}^{}(x,i\kappa_{m})}{\tau (x)}e^{K_{m}(x)},\qquad \Psi_{\kappa_{m}}(x)=\dfrac{\displaystyle\res_{\bk=i\kappa_{m}}\tau_{\Psi}^{}(x,\bk)}{\tau (x)}e^{-K_{m}(x)},
\label{discr}
\end{equation}
where $m=1,\ldots ,\No$. We consider the asymptotic of $\Phi (x,i\kappa_{m})$, since the behavior of $\Psi_{\kappa_{m}}(x)$ can be obtained from (\ref{-PhiPsi}) that for these discrete values means
\begin{equation}
\Psi_{\kappa_{m}}(x)=i(-1)^{\No}\gamma_{m}\bigl[\Phi(-x,i\kappa_{m})\bigr],\label{PPn}
\end{equation}
where we used notation (\ref{gamma}) and where the square brackets denotes operation (\ref{PP}).

Let us fix some $n=1,\ldots ,\No$ and let $x\overset{\overline{\sigma^{\pa}_{n}}}{\longrightarrow}\infty$. From (\ref{Phik1}) we have
\begin{equation}\label{asympt1}
\Phi (x,i\kappa_{m})=\left\{\begin{array}{cl}
a_{m}^{(n)}(x), & m=n,\ldots ,n+N_{b}-1, \\
p_{m}^{(n)}e^{K_{m}(x)}, & m=n+N_{b},\ldots ,n+\No-1,\end{array}\right. +\ldots ,
\end{equation}
where constants $p_{m}^{(n)}$ equal
\begin{equation}
p_m^{(n)}=\prod_{j=n}^{n+N_b-1}(\kappa_{j}-\kappa_{m}),  \label{pmn}
\end{equation}
and where functions $a_{m}^{(n)}(x)$, that give the leading asymptotic behavior for the corresponding values of $m$, should be determined (thanks to (\ref{Phik1}) we know only that they tend to zero when  $x\overset{\overline{\sigma^{\pa}_{n}}}{\longrightarrow}\infty $).

Using the permutation introduced in (\ref{d62}) we write
\begin{align}
& \Phi (x,i\kappa)\pi_{n}\equiv \bigl(\Phi (x,i\kappa_{m+n+N_{b}-1})\bigr)_{m=1}^{\No}=  \notag \\
& =\Bigl(p_{n+N_{b}}^{(n)}e^{K_{n+N_{b}}(x)},\ldots ,p_{n+\No-1}^{(n)}e^{K_{n+\No-1}(x)},
a_{n}^{(n)},\ldots ,a_{n+N_{b}-1}^{(n)}\Bigr)+\ldots.  \label{asympt3}
\end{align}
Now by (\ref{blockzn}) we get from (\ref{d6})
\begin{equation*}
a_{m}^{(n)}=-\sum_{l=1}^{N_{a}}p_{n+l+N_{b}-1}^{(n)}e^{K_{n+l+N_{b}-1}(x)}(d_{(n)})_{l,m-n+1},
\quad m=n,\ldots ,n+N_{b}-1,
\end{equation*}
where we used that matrices $v_{n}$ are invertible. Inserting these relation in (\ref{asympt3}) and again using (\ref{blockzn}) we get
\begin{align}
& \Phi (x,i\kappa)=\Bigl(p_{n+N_{b}}^{(n)}e^{K_{n+N_{b}}(x)},\ldots ,p_{n+\No-1}^{(n)}
e^{K_{n+\No-1}(x)}\Bigr){v_{n}'}^{-1}\Do^{\,\prime}+\ldots ,\label{asympt5} \\
& \Psi_{\kappa}(x)=-i\Do{v_{n}}^{-1}\Bigl(\bigl(p_{n}^{(n)}\bigr)^{-1}
e^{-K_{n}(x)},\ldots ,\bigl(p_{n+N_{b}-1}^{(n)}\bigr)^{-1}e^{-K_{n+N_{b}-1}(x)}\Bigr)^{\text{T}}+\ldots,  \label{asympt6}
\end{align}
where the second equality is derived by means of (\ref{PPn}) and (\ref{blockzn}), (\ref{pmn}). Thanks to (\ref{d55}) this means that for $x\overset{\overline{\sigma^{\pa}_{n}}}{\longrightarrow}\infty $ we proved that
\begin{align}
& \varphi (x)=\Bigl(p_{n+N_{b}}^{(n)}e^{K_{n+N_{b}}(x)},\ldots ,p_{n+\No-1}^{(n)}
e^{K_{n+\No-1}(x)}\Bigr){v_{n}'}^{-1}+\ldots ,\label{asympt7} \\
& \psi (x)=-i{v_{n}}^{-1}\Bigl(\bigl(p_{n}^{(n)}\bigr)^{-1}e^{-K_{n}(x)},\ldots ,\bigl(p_{n+N_{b}-1}^{(n)}\bigr)^{-1}
e^{-K_{n+N_{b}-1}(x)}\Bigr)^{\text{T}}+\ldots .  \label{asympt8}
\end{align}
Asymptotic values of $\varphi (x)$ and $\psi (x)$ depend on the choice of matrices $\Do$ and $\Do'$ (cf. discussion
after (\ref{d62})), while asymptotics (\ref{asympt5}) and (\ref{asympt6}) are invariant. In fact we can give a more detailed description of this asymptotic behavior. For this sake let us divide sectors $\sigma_{n}$ in two subsectors
\begin{equation}
\begin{split}
& \sigma_{n}'=\{x:K_{n+N_{b}-1}(x)>K_{n-1}(x)\text{ and } K_{n-1}(x)>K_{n+N_{b}}(x)\}, \\
& \sigma_{n}^{\prime \prime}=\{x:K_{n+N_{b}}(x)>K_{n-1}(x)\text{ and }K_{n}(x)>K_{n+N_{b}}(x)\}.
\end{split}\label{f20}
\end{equation}
It is easy to see that
\begin{equation}
\sigma_{n}=\sigma_{n}'\cup \sigma_{n}^{\prime \prime}\cup {r_{n}'},  \label{f21}
\end{equation}
where $r_{n}'$ is the ray belonging to the sector $\sigma_n$
\begin{equation}
r_{n}'=\{x:K_{n-1}(x)=K_{n+N_{b}}(x)\text{ and }(\kappa_{n+N_{b}}-\kappa_{n-1})x_{2}<0\}.  \label{rn}
\end{equation}
In analogy to (\ref{pointn'}) we can determine it by means of the directing vector
\begin{equation}
y_{n}'=(\kappa_{n+N_{b}}^{2}-\kappa_{n-1}^{2},\kappa_{n-1}-\kappa_{n+N_{b}}).  \label{rn'}
\end{equation}
In the case $N_{a}=1$, i.e., $N_{b}=\No-1$ by~(\ref{Nnanb}), these definitions are senseless as it follows from~(\ref{modN}). So we will consider this case below separately. Now in analogy to Definition~\ref{def1'} we introduce
\begin{definition}
\label{def2} We say that $x\to\infty$ in the closed subsector $\overline{\sigma'_{n}}$ (notation $x\overset{\overline{\sigma'_{n}}}{\longrightarrow}\infty$), or in the closed subsector
$\overline{\sigma^{\prime \prime}_{n}}$
(notation $x\overset{\overline{\sigma^{\prime \prime}_{n}}}{\longrightarrow}\infty$), $n=1,\ldots\No$, if there exist nonnegative $\alpha$ and $\beta$ such that $\alpha+\beta\to+\infty$ and
\begin{align*}
&\text{for}&&x\overset{\overline{\sigma'_{n}}}{\longrightarrow}\infty: & &x-\alpha
y_{n-1}-\beta y'_{n}\text{ is bounded,}\\
&\text{for}&&x\overset{\sigma^{\prime \prime}_{n}}{\longrightarrow}\infty: & &x-\alpha
y'_{n}-\beta y_{n}\text{ is bounded.}
\end{align*}
\end{definition}
Then we have the following lemma.
\begin{lemma}
\label{lemma-add}
\begin{enumerate}
\item If $x\overset{\overline{\sigma_{n}'}}{\longrightarrow}\infty $ then $K_{n-1}(x)>K_{m}(x)$ for
any $m=n+N_{b},\ldots ,n+\No-2$ and $K_{m}(x)>K_{n+N_{b}-1}(x)$ for any $m=n,\ldots ,n+N_{b}-2$;
\item if $x\overset{\overline{\sigma_{n}^{\prime \prime}}}{\longrightarrow}\infty$ then $K_{n+N_{b}}(x)>K_{m}(x)$ for any $m=n+N_{b}+1,\ldots ,n+\No-1$ and $K_{n}(x)>K_{m}(x)$ for any $m=n+1,\ldots ,n+N_{b}-1$.
\end{enumerate}
\end{lemma}

\textsl{Proof.\/} Thanks to Definition \ref{def2}:
$K_{n-1}(x)-K_{m}(x)=\alpha g_{n-1,m,n-1}+\beta\widetilde{g}_{n-1,m,n-1}$ and
$K_{m}(x)-K_{n+N_{b}-1}(x)=\alpha g_{m,n-1,n-1}+\beta\widetilde{g}_{m,n+N_{b}-1,n-1}$, where $g_{l,m,n}$ is defined in (\ref{l11}) and $\widetilde{g}_{l,m,n}$ is $g_{l,m,n}$ with substitution $N_b\to{N_b+1}$. Then the first
statement follows by lemma \ref{lemma1}. In the same way we get
$K_{n+N_b}(x)-K_{m}(x)=\alpha\widetilde{g}_{n-1,m,n-1}+\beta g_{n,m,n}$ and
$K_{m}(x)-K_{n}(x)=\alpha \widetilde{g}_{m,n,n-1}+\beta g_{m,n,n}$ and the second statement results again from Lemma \ref{lemma1}.$\blacksquare$

Thanks to this Lemma we get from (\ref{asympt5}):
\begin{align}
& \Phi (x,i\kappa_{m})=p_{n-1}^{(n)}\Bigl({v_{n}'}^{-1}\Do'\bigr)^{}_{N_a,m}e^{K_{n-1}(x)}+\ldots,\quad x\overset{\overline{\sigma_{n}'}}{\longrightarrow}\infty ,\label{asympt9} \\
& \Phi (x,i\kappa_{m})=p_{n+N_{b}}^{(n)}\Bigl({v_{n}'}^{-1}\Do^{\prime}\Bigr)^{}_{1,m}e^{K_{n+N_{b}}(x)}+\ldots ,\quad x\overset{\overline{\sigma_{n}^{\prime \prime}}}{\longrightarrow}\infty,  \label{asympt10}
\end{align}
for any $n=1,\ldots,\No$ and $m=n,\dots ,n+N_{b}-1$. In the exceptional case $N_{a}=1$ it is easy to see that thanks to (\ref{modN}) both these equalities coincide, and we can use any of them for the asymptotic behavior in the whole sector $\overline{\sigma^{\pa}_{n}}$.  Correspondingly, relations for the asymptotic values of $\Psi_{\kappa_{m}}(x)$ follow from (\ref{PPn}) and (\ref{asympt6}), where it is necessary to take into account that images of subsectors (\ref{f20}) are different from $\sigma'_{n+N_a}$ and $\sigma''_{n+N_a}$, cf. (\ref{sigmasigma}). Exactly, we get decomposition of sectors $\sigma_n$ different from (\ref{f21}):
\begin{equation}
\overline{\sigma^{\phantom{a}}_{n}}=\overline{\bigl[\sigma'_{n+N_a}\bigr]\cup\bigl[\sigma''_{n+N_a}\bigr]},
\label{saigma:br}
\end{equation}
where square brackets denotes by (\ref{PP}) substitution $N_{a}\leftrightarrow N_{b}$. So that by (\ref{f20})
\begin{equation}
\begin{split}
& [\sigma_{n+N_a}']=\{x:K_{n+N_{b}-1}(x)>K_{n-1}(x)\text{ and } K_{n}(x)>K_{n+N_{b}-1}(x)\}, \\
& [\sigma_{n}'']=\{x:K_{n+N_{b}-1}(x)>K_{n}(x)\text{ and }K_{n}(x)>K_{n+N_{b}}(x)\}.
\end{split}\label{f22}
\end{equation}
The same remark as above must be given in the case $N_b=1$.

In the following we need the asymptotics of $\Phi (x,i\kappa_{m})$ and $\Psi_{\kappa_{m}}(x)$ for fixed $m$ in dependence on $n$, while in formulas above we have its asymptotic for fixed $n$ in dependence of running $m$. These asymptotics can be, easily, obtained from  (\ref{asympt9}), (\ref{asympt10}) and analogous formulas for $\Psi_{\kappa_{m}}(x)$, and we can state the following theorem.

\begin{theorem}
\label{th3} The leading asymptotic behavior of the discrete values of the Jost solutions $\Phi (x,\bk)$  and $\Psi(x,\bk)$ is given by
\begin{align}
&\Phi (x,i\kappa_{m})=\nonumber\\
&=\left\{\begin{array}{lll}
p_{m}^{(n)}e^{K_{m}(x)}, & x\overset{\overline{\sigma^{\pa}_{n}}}{\longrightarrow}\infty , & n\in[m+1,m+N_{a}], \\
p_{n-1}^{(n)}\left( {v_{n}'}^{-1}\Do'\right)_{N_{a},m}e^{K_{n-1}(x)}, &
x\overset{\overline{\sigma_{n}'}}{\longrightarrow}\infty , & n\in[m+N_{a}+1,m+\No], \\
p_{n+N_{b}}^{(n)}\left( {v_{n}'}^{-1}\Do'\right)_{1,m}e^{K_{n+N_{b}}(x)}, &
x\overset{\overline{\sigma_{n}^{\prime \prime}}}{\longrightarrow}\infty , & n\in[m+N_{a}+1,m+\No],\end{array}
\right. \label{asympt-jost1}\\
&\qquad\qquad+\ldots\nonumber\\
&i\Psi_{\kappa_{m}}(x)=\nonumber\\
&=\left\{\begin{array}{lll}
(p_{m}^{(n)})^{-1}e^{-K_{m}(x)}, & x\overset{\overline{\sigma^{\pa}_{n}}}{\longrightarrow}\infty , & n\in[m+N_{a}+1,m+\No], \\
&&\\
\dfrac{\left( \Do{v_{n}}^{-1}\right)_{mN_{b}}}{p_{n+N_{b}-1}^{(n)}}e^{-K_{n+N_{b}-1}(x)}, & x\overset{\overline{[\sigma_{n+N_a}']}}{\longrightarrow}\infty , & n\in[m+1,m+N_{a}], \\
&&\\
\dfrac{\left( \Do{v_{n}}^{-1}\right)_{m1}}{p_{n-1}^{(n)}}e^{-K_{n}(x)}, & x\overset{\overline{[\sigma''_{n+N_a}]}{\longrightarrow}}\infty , & n\in[m+1,m+N_{a}],
\end{array}\right.\label{asympt-jost2}\\
&\qquad\qquad+\ldots\nonumber
\end{align}
\end{theorem}

We see that, for $n=m+1,\ldots ,m+N_{a}$, the Jost solution $\Phi (x,i\kappa_{m})$ has the same asymptotics along the directions on the $x$-plane obtained by considering the union of the corresponding sectors $\overline{\sigma^{\pa}_{n}}$, that is, thanks to~(\ref{sigman}), along the following directions
\begin{align}
& \bigcup_{n=m+1}^{m+N_{a}}\overline{\sigma_{n}^{}}=  \label{f9} \\
& =\left\{\begin{array}{ll}
\{K_{m+N_{a}}(x)-K_{m}(x)>-\infty \}\cap \{K_{m+N_{b}}(x)-K_{m}(x)>-\infty\}, & N_{a}<N_{b}, \\
\{K_{m+N_{a}}(x)-K_{m}(x)>-\infty \}, & N_{a}=N_{b}, \\
\Rs^{2}\setminus \{K_{m}(x)-K_{m+N_{a}}(x)>-\infty \}\cap\{K_{m}(x)-K_{m+N_{b}}(x)>-\infty \}, & N_{a}>N_{b},
\end{array}\right.  \notag
\end{align}
where, say, condition $K_{m+N_{a}}(x)-K_{m}(x)>-\infty $ means that $x\rightarrow \infty $ in such way that this difference is bounded from below. Geometrically the set in (\ref{f9}) essentially depends on relation between $N_{a}$ and $N_{b}$. Thus we have, in the first line, a sector with a sharp angle at the vertex (being the intersection of two halfplanes), in the second line, a halfplane and, in the third line, a sector with a blunt angle at the vertex.

\section{Annihilators}

In the theory of the one-dimensional Sturm--Liouville operator it is well known that the discrete value of the Jost solution corresponding to the one soliton potential (i.e., the potential in~(\ref{1-sol}) with $\kappa_{1}=-\kappa_{2}$) decays exponentially on the $x_{1}$-axis as $e^{-|x_{1}\kappa_{1}|}$. The Jost solution of the heat equation, obviously, cannot obey analogous property on the whole $x$-plane because of the $x_{2}$-dependence of the exponential factor (see, e.g.,~(\ref{discr})).

\begin{figure}[tbp]
\begin{center}
\begin{pspicture}[linewidth=1.0pt](-4,-1)(4,7)
\psset{xunit=0.7cm,yunit=0.7cm,runit=0.7cm}
\parabola(-4,6)(0,0)\rput[lB]{-70.5}(-3.73,5.9){${q_2=q_1^2}$}
\psline[linecolor=black]{->}(0,-0.6)(0,6.3)
\psline[linecolor=black]{->}(-4,0)(4.2,0)
\rput(4.3,0.15){$q_1$}
\rput(-0.2,6){$q_2$}
\rput(-3,-0.3){$\kappa_{1}$}\rput(2.6,-0.3){$\kappa_{2}$}
\psdots(-3,0)(2.6,0)
\psdots(-3,3.4)(2.6,2.56)
\psline[linestyle=dashed](-3,0)(-3,3.4)\psline[linestyle=dashed](2.6,0)(2.6,2.6)
\psline[linecolor=black](-4,3.55)(-3,3.4)
\psline[linecolor=yellow](-3,3.4)(2.6,2.6)
\psline[linecolor=black](2.6,2.6)(4,2.38)
\end{pspicture}
\vskip0.2cm
\end{center}
\caption{One-dimensional case}
\label{fig1}
\end{figure}
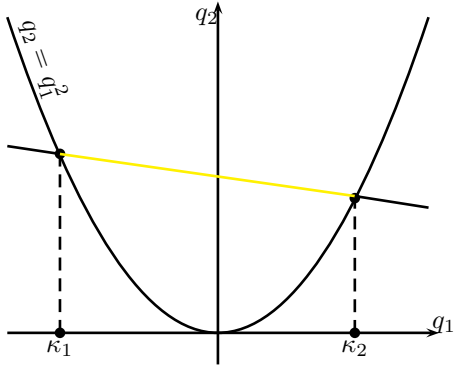

In order to find a hint for a proper formulation in the case of the heat equation, let us consider the trivial example of the one soliton potential, i.e., the case $N_{a}=N_{b}=1$ and arbitrary $\kappa_{1}$ and $\kappa_{2}$ obeying~(\ref{kappas}). By~(\ref{tauf}) and (\ref{f}) we have $\tau(x)=de^{K_{1}(x)}+e^{K_{2}(x)}$, where $d$ is a constant, that thanks to~(\ref{ux}) gives~(\ref{1-sol}). Then from~(\ref{discr}) we have
\begin{equation}
\Phi(x,i\kappa_{2})=-d\Phi(x,i\kappa_{1})=\dfrac{d(\kappa_{1}-\kappa_{2})e^{K_{1}(x)+K_{2}(x)}}{de^{K_{1}(x)}+
e^{K_{2}(x)}},  \label{1sol:jost}
\end{equation}
and for the asymptotics of the discrete value of the Jost solution
\begin{equation}
\Phi (x,i\kappa_{1})\cong \Phi (x,i\kappa_{2})\cong \left\{\begin{array}{ll}e^{K_{2}(x)}, & x\overset{\overline{\sigma_{1}}}{\longrightarrow}\infty ,\\
e^{K_{1}(x)}, & x\overset{\overline{\sigma_{2}}}{\longrightarrow}\infty ,\end{array}\right.  \label{asympt-jost1sol}
\end{equation}
where closed sectors $\overline{\sigma^{\pa}_{1}}$ and $\overline{\sigma^{\pa}_{2}}$ are given by
\begin{equation}\begin{split}
& \overline{\sigma_{1}}=\{K_{1}(x)-K_{2}(x)>-\infty \}\equiv\{x_{1}+(\kappa_{1}+\kappa_{2})x_{2}<+\infty \}, \\
& \overline{\sigma_{2}}=\{K_{2}(x)-K_{1}(x)>-\infty \}\equiv\{x_{1}+(\kappa_{1}+\kappa_{2})x_{2}>-\infty \},
\end{split}
\end{equation}
where we understand inequalities here like in definition (\ref{def3}). Let, now, denote the scalar product of the two-dimensional vectors $x=(x_{1},x_{2})$ and $q=(q_{1},q_{2})$ as
\begin{equation}
qx=q_{1}x_{1}+q_{2}x_{2},  \label{qx}
\end{equation}
and let
\begin{equation}
q_{mn}=q_{2}-(\kappa_{m}+\kappa_{n})q_{1}+\kappa_{m}\kappa_{n}\label{qmn}
\end{equation}
for any $m,n\in \Zs$. Then, thanks to~(\ref{asympt-jost1sol}), we get for the asymptotic behavior of the functions
$e^{-qx}\Phi (x,i\kappa_{1})$ and $e^{-qx}\Phi (x,i\kappa_{2})$
\begin{align}
e^{-qx}\Phi (x,i\kappa_{1})& \cong e^{-qx}\Phi (x,i\kappa_{2})\cong  \notag\\
& \cong \left\{\begin{array}{ll}
\exp [-(q_{1}-\kappa_{2})(x_{1}+(\kappa_{1}+\kappa_{2})x_{2}-q_{12}x_{2}],
& x\overset{\overline{\sigma_{1}}}{\longrightarrow}\infty , \\
\exp [-(q_{1}-\kappa_{1})(x_{1}+(\kappa_{1}+\kappa_{2})x_{2}-q_{12}x_{2}],
& x\overset{\overline{\sigma_{2}}}{\longrightarrow}\infty ,\end{array}\right.  \label{1-solit}
\end{align}
and we deduce that they are bounded for all $x$ iff $q_{12}=0$ and $\kappa_{1}<q_{1}<\kappa_{2}$, this on the segment of the line $q_{12}=0$ on the $q$-plane cut off by the parabola $q_{2}=q_{1}^{2}$ (see Fig.~\ref{fig1}).

This trivial result has unexpected generalization in the case $N_{a}\neq{N_{b}}$. Precisely, for $N_{b}>N_{a}\geq1$, the set of the $q$-plane, where the discrete values $\Phi (x,i\kappa_{m})$ of the Jost solution multiplied by the exponent $e^{-qx}$ are bounded, becomes a polygon. As regards the dual Jost solution, the functions $e^{qx}\Psi_{\kappa_{m}}(x)$ are bounded
inside a polygon for $N_{a}>N_{b}\geq1$.

Let, then, consider the case $N_{b}>N_{a}\geq1$ and let us introduce the angular sectors
\begin{equation}
\rho_{k}(q)=\theta (\widetilde{q}_{k,k-N_{a}})\theta (\widetilde{q}_{k+N_{a},k})  \label{rhok}
\end{equation}
where the $\theta$'s are the step functions and
\begin{equation}
\widetilde{q}_{mn}=(\kappa_{m}-\kappa_{n})q_{mn},  \label{f31}
\end{equation}
and let us call
\begin{equation}
Q_{m}=(\kappa_{m},\kappa_{m}^{2}),\qquad Q_{n}=(\kappa_{n},\kappa_{n}^{2}),
\end{equation}
the points of intersection of the line $\widetilde{q}_{mn}=0$ with the parabola $q_{2}=q_{1}^{2}$ in the $q$-plane.

The angular sector $\rho_{k}(q)$ has vertex on the parabola at the point $Q_{k}$. The ray bordering the sector from the left (looking from the vertex inside the sector) is crossing the parabola at the point $Q_{k-N_{a}}$ and the ray bordering the sector from the right (looking from the vertex inside the sector) is crossing the parabola at the point $Q_{k+N_{a}}$. If we add to the parabola the point at infinity and, then, we consider it as a closed curve, as $k$ increases the sector move along the parabola in the anticlockwise direction.

Then, we can prove the following Lemmas.

\begin{lemma}
\label{inside} The region of the $q$-plane defined by the characteristic function
\begin{equation}
P_{m}(q)=\prod_{k=m+N_{a}}^{m+N_{b}}\rho_{k}(q)  \label{Pm}
\end{equation}
is a polygon included in the parabolic region $q_{2}\geq q_{1}^{2}$ with a vertex at the point $Q_{m}$. More precisely this polygon is included in the polygon with vertices the points $Q_{m}$, $Q_{m+1}$, $\dots $, $Q_{m+N_{b}}$ belonging to the parabola, coincides with this polygon in the case $N_{a}=1$ and belongs to the strip
\begin{equation}
\min_{m\leq l\leq m+N_{b}}\kappa_{l}\leq {q_{1}}\leq \max_{m\leq l\leq m+N_{b}}\kappa_{l}.  \label{strp}
\end{equation}
\end{lemma}

\textsl{Proof.\/} The angular sector $\rho_{k}$ has at $k=m+N_{a}$ the left (looking from the vertex inside the sector) ray crossing the parabola at the point $Q_{m}$. Increasing $k$ the sector rotates along the parabola in the anticlockwise direction up to the sector $\rho_{m+N_{a}}$, that has right (looking from the vertex inside the sector) ray crossing the parabola just at the point $Q_{m}$. Therefore the sectors $\rho_{k}$ for $k$ running in the interval $k=m+N_{a},\dots m+N_{b}$ cover a common region, which is a polygon inside the parabola with a vertex at the point $Q_{m}$. For $N_{a}=1$ each sector $\rho_{k}$ has right ray coinciding with the left ray of the sector $\rho_{k+1}$ and, therefore, the sides of the polygon are just the intersections of the rays bordering the angular sectors $\rho_{k}$ with the region inside the parabola and the vertices of the polygon are the points $Q_{m}$, $Q_{m+1}$, $\dots $, $Q_{m+N_{b}}$ belonging to the parabola. For $N_{a}>1$ the point $Q_{k+N_{a}}$ intersection of the right ray of the sector $\rho_{k}$ with the parabola lies to the left of the left ray of the sector $\rho_{k+1}$ and, therefore, does not belong to the polygon and we deduce that the polygon has only one vertex belonging to the parabola and, precisely, the point $Q_{m}$ and that it belongs to the polygon with vertices the points $Q_{m}$, $Q_{m+1}$, $\dots $, $Q_{m+N_{b}}$. In addition, we can deduce that the polygon defined by (\ref{Pm}) belongs to the strip in (\ref{strp}).$\blacksquare$

\begin{lemma}
\label{l52} The polygon defined by the characteristic function
\begin{equation}
\epsilon_{m}(q)=\prod_{k=m}^{m+N_{b}}\theta (\widetilde{q}_{k+N_{a},k})\label{epsilonm}
\end{equation}
satisfies the equation
\begin{equation}
\epsilon_{m}(q)=P_{m}(q)\left\{\begin{array}{ll}
1 & N_{b}\geq 2N_{a}-1 \\
\displaystyle\prod_{k=m+N_{b}-N_{a}+1}^{m+N_{a}-1}\theta (\widetilde{q}_{k+N_{a},k}), & N_{b}<2N_{a}-1.
\end{array}\right.,  \label{f33}
\end{equation}
and, therefore, belongs to the polygon $P_{m}(q)$ considered in Lemma \ref{inside} and is included in the parabolic region $q_{2}\geq q_{1}^{2}$ of the $q$-plane. Moreover, the polygon defined by (\ref{epsilonm}) is not void.
\end{lemma}

\textsl{Proof.\/} Thanks to the equality $\prod_{k=m}^{m+N_{b}}\theta(\widetilde{q}_{k+N_{a},k})=\prod_{k=m+N_{a}}^{m+\No}
\theta (\widetilde{q}_{k,k-N_{a}})$ obtained by shifting $k\rightarrow k-N_{a}$, we get that
\begin{equation*}
\prod_{k=m}^{m+N_{b}}\theta (\widetilde{q}_{k+N_{a},k})=\left(\prod_{k=m}^{m+N_{b}}\theta (\widetilde{q}_{k+N_{a},k})\right)\prod_{k=m+N_{a}}^{m+\No}\theta (\widetilde{q}_{k,k-N_{a}}),
\end{equation*}
as $\theta $'s of the second factor are all present in the first one. Then, factorizing $P_{m}(q)$ from the r.h.s.\ and shifting the running index $k$ for the remaining $\theta $'s in the two products for $k\rightarrow k-N_{a}$ and $k\rightarrow k+N_{a}$, respectively, we get
\begin{equation}
\epsilon_{m}(q)=P_{m}(q)\left( \prod_{k=m+N_{a}}^{m+2N_{a}-1} \theta(\widetilde{q}_{k,k-N_{a}})\right) \prod_{k=m+N_{b}-N_{a}+1}^{m+N_{b}} \theta(\widetilde{q}_{k+N_{a},k}).
\end{equation}
If $N_{b}\geq 2N_{a}-1$, the $\theta $'s in the two products in the r.h.s\ are already present in $P_{m}(q)$ and, then we get the first equality in (\ref{f33}). If $N_{b}<2N_{a}-1$, we have
\begin{align}
& \prod_{k=m+N_{a}}^{m+2N_{a}-1}\theta (\widetilde{q}_{k,k-N_{a}})=\prod_{k=m+N_{a}}^{m+N_{b}}
\theta (\widetilde{q}_{k,k-N_{a}})\prod_{k=m+N_{b}+1}^{m+2N_{a}-1}\theta (\widetilde{q}_{k,k-N_{a}})\label{101} \\
& \prod_{k=m+N_{b}-N_{a}+1}^{m+N_{b}-1}
\theta (\widetilde{q}_{k+N_{a},k})=\left(\prod_{k=m+N_{b}-N_{a}+1}^{m+N_{a}-1}\theta (\widetilde{q}_{k+N_{a},k})\right) \prod_{k=m+N_{a}}^{m+N_{b}}\theta (\widetilde{q}_{k+N_{a},k}),  \label{102}
\end{align}
and, then, recalling the expression for $P_{m}(q)$ in (\ref{Pm}), (\ref{rhok}) and noticing that the second product in (\ref{101}) and the first product in (\ref{102}) coincide, we get the second equality in (\ref{f33}).

In order to prove that the polygon defined by (\ref{epsilonm}) is not void it is sufficient to prove that the product in the second line of the r.h.s.\ of (\ref{f33}) equals 1 at the point $Q_{m}$. In fact we have from definitions (\ref{l11}) and (\ref{f31}) that $\widetilde{q}_{k+N_{a},k}(Q_{m})=-g_{k+N_{a},m,k+N_{a}}$, which, thanks to Lemma \ref{lemma1}, is greater or equal to zero only for $m=k,\dots ,k+N_{a}$ (mod$\No$), which is impossible for $k=m+N_{b}-N_{a}+1,\dots ,m+N_{a}-1$ and $N_{b}<2N_{a}-1$. $\blacksquare$

\begin{lemma}
\label{l53} The characteristic functions $\epsilon_{m}(q)$ introduced in Lemma \ref{l52} coincide with the characteristic function.
\begin{equation}
\widetilde{\epsilon}_{m}(q)=\prod_{k=m}^{m+N_{b}}\theta (\widetilde{q}_{k+N_{a},k})\prod_{k=m}^{m+N_{b}-1}
\theta (\widetilde{q}_{l+N_{a},l+1}),
\end{equation}
\end{lemma}

\textsl{Proof.\/} It follows if we notice that the parameters $\kappa_{N_{a}+1}$, $\kappa_{l}$, and $\kappa_{l+1}$ for any value of $l$ result to be ordered only in one of the two following ways: $\kappa_{N_{a}+1}<\kappa_{l}<\kappa_{l+1}$ or $\kappa_{l}<\kappa_{l+1}<\kappa_{N_{a}+1}$ with the only exception of the case $l=\No$ (mod$\,\No$), where $\kappa_{\mathcal{N}+1}<\kappa_{N_{a}+1}<\kappa_{\mathcal{N}}$. Taking into account that $q$ is inside the parabola $q_{2}=q_{1}^{2}$, we get for any value of $l$ that
$\theta (\widetilde{q}_{l+N_{a},l})\theta (\widetilde{q}_{l+N_{a},l+1})=\theta (\widetilde{q}_{l+N_{a},l})$. $\blacksquare $

Now, we can prove the following Theorem.

\begin{theorem}\label{th2}

\begin{enumerate}
\item\label{th2-1}  In the case $N_{a}<N_{b}$, the function $e^{-qx}\Phi (x,i\kappa_{m})$, for any $m=1,\dots ,\mathcal{N}$, is exponentially decreasing with respect to $x$ for any $q$ inside the polygon defined by the characteristic function
\begin{equation}
\varepsilon_{m}(q)=\prod_{k=m}^{m+N_{b}}\theta (\widetilde{q}_{k+N_{a},k}),\label{f41}
\end{equation}
while for $N_{a}\geq N_{b}$ there exists no such domain on the $q$-plane.
\item\label{th2-2}  In the case  $N_{a}>N_{b}$  the function $e^{qx}\Psi_{\kappa_{m}}(x)$, for any $m=1,\dots ,\mathcal{N}$, is exponentially decreasing with respect to $x$ for any $q$ inside the polygon defined by the characteristic function
\begin{equation}
\varepsilon_{m}(q)=\prod_{k=m}^{m+N_{b}}\theta (\widetilde{q}_{k,k+N_{a}}),\label{f41'}
\end{equation}
while for  $N_{a}\leq N_{b}$ there exists no such domain on the $q$-plane.
\item These polygons are all included in the parabolic region $q_{2}\geq q_{1}^{2}$.
\end{enumerate}
\end{theorem}

\textsl{Proof.\/} In (\ref{asympt-jost1}) and (\ref{f9}) we have seen that the geometry of the sector $\overline{\Sigma^{\phantom{f}}_{m}}=\bigcup_{n=m+1}^{m+N_{a}}\overline{\sigma_{n}^{\phantom{a}}}$ where $\Phi (x,i\kappa_{m})$ has asymptotic behavior $e^{K_{m}(x)}$ essentially depends on the relation between $N_{a}$ and $N_{b}$. Then, first, let $N_{a}<N_{b}$ and let us consider the directing vectors of the rays bordering the sectors $\overline{\Sigma^{\phantom{f}}_{m}}$, $\overline{\sigma_{n}^{\prime}}$ and $\overline{\sigma_{n}^{\prime\prime}}$ ($n=m+N_{a}+1,\dots ,m+\mathcal{N}$), where the product $e^{-qx}\Phi (x,i\kappa_{m})$, according to (\ref{asympt-jost1}) and (\ref{f9}), has, respectively, asymptotic behavior $e^{K_{m}(x)-qx}$, $e^{K_{n-1}(x)-qx}$ and $e^{K_{n+N_{b}}(x)-qx}$. For $\overline{\Sigma^{\phantom{f}}_{m}}$ they are $y_{m}$ and $y_{m+N_{a}}$, for $\overline{\sigma_{n}^{\prime}}$ they are $y_{n-1}^{}$ and $y_{n}^{\prime}$, and for $\overline{\sigma_{n}^{\prime \prime}}$ they are $y_{n}^{}$ and $y_{n}^{\prime}$ (see (\ref{pointn'}) and (\ref{rn'})). According to Definition \ref{def1'} we say that $x\rightarrow \infty $ in the sector $\overline{\Sigma^{\phantom{f}}_{m}}$ if $x-\alpha y_{m}-\beta y_{m+N_{a}}$ is bounded while $\alpha^{2}+\beta^{2}\rightarrow \infty$ and both $\alpha$ and $\beta$ are bounded from below. On the other side, it is easy to see that
\begin{equation}
K_{m}(x)-qx\bigr|_{x=\alpha y_{m}+\beta y_{m+N_a}}=\alpha \widetilde{q}_{m+N_{b},m}-\beta\widetilde{q}_{m+N_{a},m}.  \label{th31}
\end{equation}
and we deduce that $e^{-qx}\Phi (x,i\kappa_{m})$ is bounded in the sector $\overline{\Sigma^{\phantom{f}}_{m}}$ iff $\widetilde{q}_{m+N_{b},m}\leq 0$ and $\widetilde{q}_{m+N_{a},m}\geq 0$. By Definition \ref{def2} we say that $x\rightarrow \infty $ in the sectors $\overline{\sigma_{n}^{\prime}}$ and $\overline{\sigma_{n}^{\prime \prime}}$ if $x-\alpha y_{n-1}^{}-\beta y_{n}^{\prime}$ and $x-\alpha y_{n}^{\prime}-\beta y_{n}^{}$, respectively, are bounded while $\alpha ^{2}+\beta ^{2}\rightarrow \infty $ and both $\alpha $ and $\beta $ are bounded from below. Since we have
\begin{equation}
K_{n-1}(x)-qx\bigr|_{x=\alpha y_{n-1}^{}+\beta y_{n}^{\prime}}=\alpha\widetilde{q}_{n+N_{b}-1,n-1}+
\beta \widetilde{q}_{n+N_{b},n-1}.\label{th32}
\end{equation}
and
\begin{equation}
K_{n+N_{b}}(x)-qx\bigr|_{x=\alpha y_{n}^{\prime}+\beta y_{n}^{}}=\alpha \widetilde{q}_{n+N_{b},n-1}+
\beta \widetilde{q}_{n+N_{b},n},  \label{th33}
\end{equation}
we conclude that $e^{-qx}\Phi (x,i\kappa_{m})$ is bounded in the sectors $\overline{\sigma_{n}^{\prime}}$ and $\overline{\sigma_{n}^{\prime \prime}}$ iff $\widetilde{q}_{n+N_{b}-1,n-1}$, $\widetilde{q}_{n+N_{b},n-1}$ and $\widetilde{q}_{n+N_{b},n}$ are less or equal to zero.

Thus the condition that the product $e^{-qx}\Phi (x,i\kappa_{m})$ decays asymptotically in any direction on the $x$-plane is equivalent to condition that $q$ belongs to the set on the $q$-plane given by characteristic function
\begin{equation}
\widetilde{\epsilon}_{m}(q)=\prod_{k=m}^{m+N_{b}}\theta (\widetilde{q}_{k+N_{a},k})\prod_{k=l}^{m+N_{b}-1}
\theta (\widetilde{q}_{l+N_{a},l+1}),
\end{equation}
where we took into account that indices are defined by mod$\,\No$ and that $\widetilde{q}_{m,n}=-\widetilde{q}_{n,m}$ (see (\ref{qmn}) and (\ref{f31})).

Thanks to the four Lemmas above, the first part of the statement \ref{th2-1} of the Theorem for $N_{a}<N_{b}$ is proved.

In the case $N_{a}=N_{b}$ we have by (\ref{pointn'}) that $y_{m+N_a}=-y_{m}$, $m=1,\ldots,N_a$. Thus it is enough to consider the asymptotic behavior along the ray $r_{m}$ (see Definition \ref{def1'}). We choose $y_{m}$ as directing vector and by (\ref{th31}) for $\beta =0$ we have to consider limits in both directions: $\alpha\rightarrow\pm\infty$. It is clear that the r.h.s.\ of (\ref{th31}), independently of the sign of  $\widetilde{q}_{m,m+N_{b}}$, cannot decay in both these directions and this r.h.s\ is bounded only if  $\widetilde{q}_{m,m+N_{b}}=0$, that, thanks to (\ref{qmn}) and (\ref{f31}), gives a line and not a domain on the $q$-plane. Analogously, in the case $N_{b}<N_{a}$ the directing vectors of the sector in the third line of (\ref{f9}) are the same like in (\ref{th31}), but conditions on the behavior of $\alpha $ and $\beta $ are different. Together with the limit $\alpha\rightarrow+\infty$, $\beta\rightarrow+\infty $, we have to take into account sectors $\alpha\rightarrow+\infty$, $\beta\rightarrow-\infty $ and $\alpha\rightarrow-\infty $, $\beta\rightarrow+\infty$. Thus we see that also in this case it is impossible to give conditions on $\widetilde{q}_{m,m+N_{b}}$ and $\widetilde{q}_{m,m+N_{a}}$ that always make the limit of the r.h.s.\ of (\ref{th31}) to be equal to $-\infty $. This proves statement \ref{th2-1} of the theorem. Statement \ref{th2-2} for $e^{qx}\Psi_{\kappa_{m}}(x)$ follows from (\ref{PPn}). In particular, existence of the polygon only in the case $N_{b}<N_{a} $ results from above and (\ref{PP}). The third statement of the theorem is proved in Lemma \ref{l52}. $\blacksquare$

Thus Theorem \ref{th2} is proved. Using (\ref{kappas}), (\ref{modN}), and (\ref{f31}) we can rewrite (\ref{f41}) more explicitly:
\begin{align}
& \varepsilon_{m}(q)=  \notag \\
& =\left\{\begin{array}{ll}
\displaystyle\prod\limits_{k=m}^{N_{b}}\theta(q_{k,k+N_{a}})\prod\limits_{k=1}^{m}\theta (-q_{k,k+N_{b}}),
& 1\leq m\leq N_{a}, \\
\displaystyle\prod\limits_{k=m}^{N_{b}}\theta(q_{k,k+N_{a}})\displaystyle\prod\limits_{k=1}^{N_{a}} \theta(-q_{k,k+N_{b}})\displaystyle\prod\limits_{k=N_{a}+1}^{m}\theta (q_{k,k-N_{a}}), &
N_{a}+1\leq m\leq N_{b}, \\
\displaystyle\prod\limits_{k=m-N_{b}}^{N_{a}}\theta(-q_{k,k+N_{b}})\displaystyle\prod\limits_{k=1}^{m-N_{a}}
\theta (q_{k,k+N_{a}}), &
N_{b}+1\leq m\leq \No.\end{array}\right.  \label{f42}
\end{align}

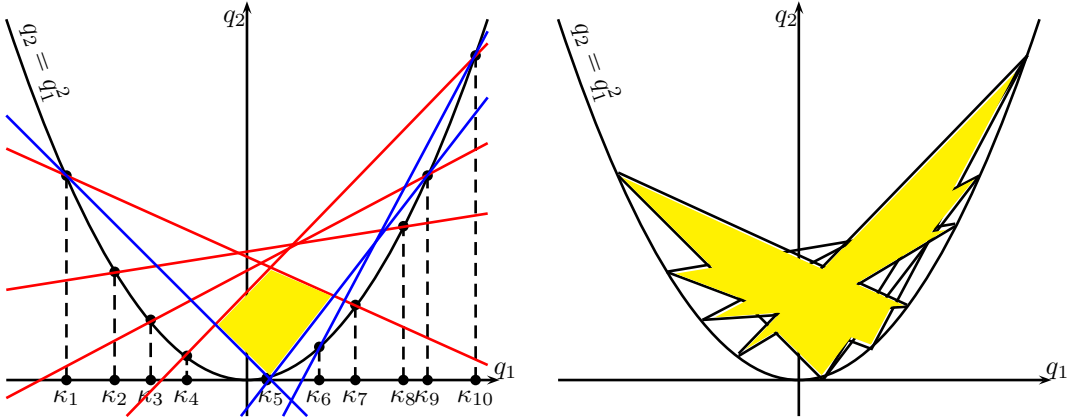
\begin{figure}[h]
\begin{pspicture}[linewidth=1.0pt](-4,-1)(4,7)
\parabola(-4,6)(0,0)\rput[lB]{-70.5}(-3.73,5.9){${q_2=q_1^2}$}
\psline[linecolor=black]{->}(0,-0.6)(0,6.3)
\psline[linecolor=black]{->}(-4,0)(4.2,0)
\rput(4.3,0.15){$q_1$}
\rput(-0.2,6){$q_2$}
\rput(-3,-0.3){$\kappa_{1}$}\rput(-2.2,-0.3){$\kappa_{2}$}\rput(-1.6,-0.3){$\kappa_{3}$}\rput(-1,-0.3){$\kappa_{4}$}
\rput(0.4,-0.3){$\kappa_{5}$}\rput(1.2,-0.3){$\kappa_{6}$}\rput(1.8,-0.3){$\kappa_{7}$}\rput(2.6,-0.3){$\kappa_{8}$}
\rput(3,-0.3){$\kappa_{9}$}\rput(3.8,-0.3){$\kappa_{10}$}
\psdots(-3,0)(-2.2,0)(-1.6,0)(-1,0)(0.33,0)(1.2,0)(1.8,0)(2.6,0)(3,0)(3.8,0)
\psdots(-3,3.4)(-2.2,1.8)(-1.6,1)(-1,0.4)(0.33,0.03)(1.2,0.55)(1.8,1.25)(2.6,2.56)(3,3.4)(3.8,5.4)
\psline[linestyle=dashed](-3,0)(-3,3.4)\psline[linestyle=dashed](-2.2,0)(-2.2,1.8)\psline[linestyle=dashed](-1.6,0)(-1.6,1)
\psline[linestyle=dashed](-1,0)(-1,0.4)\psline[linestyle=dashed](1.2,0)(1.2,0.5)\psline[linestyle=dashed](1.8,0)(1.8,1.25)
\psline[linestyle=dashed](2.6,0)(2.6,2.6)\psline[linestyle=dashed](3,0)(3,3.4)\psline[linestyle=dashed](3.8,0)(3.8,5.4)
\psline[linecolor=red](-4,3.85)(4,0.25)
\psline[linecolor=red](-4,1.5)(4,2.77)
\psline[linecolor=red](-4,-0.3)(4,3.94)
\psline[linecolor=red](-2,-0.6)(4,5.6)
\psline[linecolor=blue](-0.1,-0.6)(4,4.7)
\psline[linecolor=blue](0.6,-0.6)(4,5.8)
\psline[linecolor=blue](-4,4.4)(1,-0.6)
\pspolygon*[linecolor=yellow](-0.48,0.92)(0.38,0.08)(1.4,1.4)(0.4,1.83)
\end{pspicture}
$\qquad$
\begin{pspicture}[linewidth=1.0pt](-4,-1)(4,7)
\parabola(-4,6)(0,0)\rput[lB]{-70.5}(-3.73,5.9){${q_2=q_1^2}$}
\psline[linecolor=black]{->}(0,-0.6)(0,6.3)
\psline[linecolor=black]{->}(-4,0)(4.2,0)
\rput(4.3,0.15){$q_1$}
\rput(-0.2,6){$q_2$}
\psline(-3,3.4)(0.4,0)
\psline(-3,3.45)(1.8,1.25)
\psline(-2.2,1.8)(1.2,0.52)
\psline(-2.2,1.8)(2.6,2.6)
\psline(-1.6,1)(1.8,1.22)
\psline(-1.6,1)(3,3.4)
\psline(-1,0.4)(2.6,2.56)
\psline(-1,0.44)(3.8,5.4)
\psline(0.4,0)(3,3.4)
\psline(0.4,0)(3.8,5.4)
\psline(1.2,0.5)(3.8,5.4)
\pspolygon*[linecolor=yellow](-2.95,3.38)(-0.95,1.39)(-0.2,1.09)(0.1,1.12)(0.913,1.6)
\pspolygon*[linecolor=yellow](-2.05,1.77)(-0.3,1.11)(0.23,1.17)(0.95,1.6)(-0.14,2.08)
\pspolygon*[linecolor=yellow](-1.4,1.04)(0.21,1.16)(0.95,1.6)(0.22,1.9)
\pspolygon*[linecolor=yellow](-0.9,0.5)(0.95,1.6)(0.39,1.84)
\pspolygon*[linecolor=yellow](-0.48,0.92)(0.38,0.08)(1.4,1.4)(0.4,1.83)
\pspolygon*[linecolor=yellow](-0.28,1.14)(1.2,0.59)(1.58,1.31)(0.4,1.83)
\pspolygon*[linecolor=yellow](-0.28,1.13)(-0.2,1.1)(1.75,1.24)(0.4,1.83)
\pspolygon*[linecolor=yellow](-0.28,1.13)(-0.2,1.1)(0.18,1.15)(2.45,2.5)(0.82,2.21)(0.75,2.18)
\pspolygon*[linecolor=yellow](-0.28,1.13)(-0.2,1.1)(0.18,1.15)(2.03,2.25)(2.92,3.34)(0.72,2.18)
\pspolygon*[linecolor=yellow](-0.28,1.13)(-0.2,1.1)(0.18,1.15)(2.1,2.3)(3.6,5.12)
\end{pspicture}
\caption{$N_a=4$, $N_{b}=6$, $\No=10$, polygon for $m=5$ and union of polygons for $m=1,\dots,10$}\label{fig3}
\end{figure}

Fig.~\ref{fig3} shows, for a the potential $u(x)$ with $N_{a}=4$, $N_{b}=6$, the polygon with characteristic function $\varepsilon_{m}(q)$ on the $q$-plane in the case $m=5$ and  the union of all polygons for $m=1,\ldots,10$.

\section{Concluding remarks}

Theorem \ref{th2} shows that the extended $L$-operator
\begin{equation}
L(x,x^{\prime};q)=\{-\partial_{x_{2}^{}}^{}-q_{2}^{}+(\partial_{x_{1}^{}}^{}+q_{1}^{})_{}^{2}\} \delta (x-x^{\prime})-u(x)\delta (x-x^{\prime})  \label{L}
\end{equation}
is not invertible in the case where $u(x)$ is ($N_{a},N_{b}$)-soliton potential. If $N_{a}<N_{b}$ it has right annihilators, and in the case of $N_{b}<N_{a}$ the left ones.

In fact, in the case $N_{a}<N_{b}$, if we introduce an operator $H_{m}$ with kernel
\begin{equation}
H_{m}(x,x^{\prime};q)=\varepsilon_{m}(q)e^{-qx}\Phi(x,\kappa_{m})\psi(x^{\prime};q)
\end{equation}
where $\psi (x^{\prime};q)$ is any arbitrary self-adjoint function, bounded in $x^{\prime}$  we have
\begin{equation}
L(q) H_{m}(q)=0.
\end{equation}
Analogously, for the case $N_{b}<N_{a}$, we have
\begin{equation}
K_{m}(q)L(q)=0,
\end{equation}
with
\begin{equation}
K_{m}(x,x^{\prime};q)=\varepsilon_{m}(q)\varphi (x;q)e^{qx^{\prime}}\Psi_{\kappa_{m}}(x^{\prime}),
\end{equation}
where $\varphi (x;q)$ is any arbitrary self-adjoint function, bounded in $x$ inside the polygon defined by $\varepsilon_{m}(q)$.

We note that the total Green's function $G(x,x^{\prime},\bk)$ of the operator $\Lo$ can be defined (see~\cite{BPPPr2001a,BPPP2002}) as the value of the kernel  $M(x,x^{\prime};q)$ at $q_{1}=\bk_{\Im}$, $q_{2}=\bk_{\Im}^{2}-\bk_{\Re}^{2} $ for a complex spectral parameter $\bk=\bk_{\Re}+i\bk_{\Im}$. These values of $q$ lie outside the parabola $q_{2}=q_{1}^{2}$ and therefore outside the polygon and touch it only at the vertices. The Green's function then exists for any $\bk$ but it is singular at the points $\bk=i\kappa_{m}$ corresponding to the vertices of the polygon touching the parabola.

\section{Acknowledgments}

This work is supported in part by the grant RFBR \# 11-01-00440, Scientific Schools 8265.2010.1, by the Program of
RAS ``Mathematical Methods of the Nonlinear Dynamics,'' by INFN, by the grant PRIN 2008 ``Geometrical Methods in the theory of nonlinear integrable systems'' and by Consortium E.I.N.S.T.E.I.N. AKP thanks Department of Physics of the University of Salento (Lecce) for kind hospitality.

\appendix
\section*{Appendix}
\section{Regularity of potential and Jost solutions}
We defined the Jost solutions $\Phi (x,\bk)$ in a symmetric form in (\ref{sym}), so that
\begin{equation}
\chi (x,\bk)\equiv \Phi (x,\bk)e^{i\bk x_1+\bk^{2}x_2}
\end{equation}
became a polynomial in the spectral parameter $\bk$ with the higher power term given by $(i\bk)^{N_{b}}$. Therefore, it can be determined by giving its values in $N_{b}$ points. If these points are chosen to be $\bk=i\kappa_{n}$ with $n$ belonging to the subset $J=\{n_{1},\cdots ,n_{N_{b}}\}$ of numbers in the interval $
\{1,\ldots ,\No\}$, these values can be determined by a combined use of the analyticity properties of $\chi (x,\bk)$ and condition $\Phi (i\kappa )\Do=0$ in (\ref{d6}).

Let
\begin{equation}
\Delta (\bk)=\prod_{n\in J}(i\bk+\kappa _{n}).  \label{r1}
\end{equation}%
Then the ratio $\chi (x,\bk)\Delta ^{-1}(\bk)$ is normalized to $1$ at infinity and the poles at the points $\bk=i\kappa _{n}$ with $n\in {J}$ are the only departures from analyticity. We have, therefore,
\begin{equation}
\dfrac{\chi (x,\bk)}{\Delta (\bk)}=1+\sum_{n\in {J}}\dfrac{\chi (x,i\kappa
_{n})}{(\bk-i\kappa _{n})\Delta ^{\prime }(i\kappa _{n})}.  \label{r3}
\end{equation}%
This gives
\begin{align}
\Phi (x,i\kappa _{m})& =\Phi (x,i\kappa _{m}),\quad m\in {J},  \label{r4} \\
\Phi (x,i\kappa _{m})& =\Delta (i\kappa _{m})e^{K_{m}(x)}\left(
1-i\sum_{n\in {J}}\dfrac{\chi (x,i\kappa _{n})}{(\kappa _{m}-\kappa
_{n})\Delta ^{\prime }(i\kappa _{n})}\right) ,\quad m\notin {J},  \label{r5}
\end{align}%
Then condition $\Phi (i\kappa )\Do=0$ is equivalent to
\begin{equation}
\sum_{m\in {J}}\chi (i\kappa _{m})\left[ e^{K_{m}}\Do_{ml}-i\sum_{n\notin {J}%
}\dfrac{\Delta (i\kappa _{n})e^{K_{n}}}{(\kappa _{n}-\kappa _{m})\Delta
^{\prime }(i\kappa _{m})}\Do_{nl}\right] =-\sum_{n\notin {J}}\Delta (i\kappa
_{n})e^{K_{n}}\Do_{nl}.  \label{r6}
\end{equation}%
Comparing with (\ref{discr}) we get for the $\tau $-function (up to the standard similarity property)
\begin{equation}
\tau (x)=\det \left( \Do_{ml}-i\sum_{n\notin {J}}\dfrac{\Delta (i\kappa
_{n})e^{K_{n}-K_{m}}}{(\kappa _{n}-\kappa _{m})\Delta ^{\prime }(i\kappa
_{m})}\Do_{nl}\right) _{\substack{m\in {J}\\{l=1}{\ldots ,N_{b}}}}.  \label{r7}
\end{equation}
By inserting this $\tau $-function into (\ref{ux}) we get the soliton solution and different choices of the set $J$ correspond to different equivalent formulation of the soliton solution, in accordance with the result presented in \cite{equivKPII}.

We want to get a condition of regularity for the potential equivalent to (\ref{reg}), but involving only the $N_{a}\times N_{b}$ submatrix $d_{(n)}$ introduced in (\ref{blockzn}) and not the full $\No\times N_{b}$ matrix $\Do$.

This, of course, can be obtained by pure algebraic methods. However, we want to get this result by using the freedom in expressing the $\tau(x)$ function offered by (\ref{r7}), showing in this way its usefulness.

To the end of getting this result we consider the evolution with respect to a multivalue time describing the evolution of the entire KPII hierarchy. Then, the exponents in the different formulations of (\ref{r7}) are independent and the requirement that their coefficients are grater or equal to zero in one formulation implies the same requirement in all other formulations.

We can, then, state the following theorem.
\begin{theorem}
Let $\mathcal{D}$ be a $\No\times N_{b}$ matrix of the form
\begin{equation}
\Do=\pi \mathcal{B},\qquad \text{with\ } \mathcal{B}=\left(
\begin{array}{l}
d \\
E_{b}
\end{array}
\right) ,  \label{r33}
\end{equation}
where $d$ is a real $N_{a}\times N_{b}$ matrix and $E_{b}$ is the unit $
N_{b}\times N_{b}$ matrix and where the $\No\times \No$ matrix
\begin{equation}
\pi =\Vert \pi _{mn}\Vert _{m,n=1}^{\No},\qquad \pi _{mn}=\delta
_{\pi (m),n}\equiv \delta _{m,\pi ^{-1}(n)},  \label{r34}
\end{equation}
performs a permutation of the $\No$ rows of the matrix on the right
according to the transformation
\begin{equation}
\pi :\quad (1,\ldots ,\No)\rightarrow (\pi (1),\ldots ,\pi
(\No).  \label{r35}
\end{equation}
Then the matrix $\Do$ has all its maximal minors non negative if and only if
the $N_{a}\times N_{b}$ matrix
\begin{equation}
\widetilde{d}_{kr}=(-1)^{1+l_{r}}d_{\pi (j_{(N_{a}+1-k)}),l_{r}},\qquad
k=1,\dots ,N_{a},\quad r=1,\dots ,N_{b},  \label{r36}
\end{equation}
where $\{j_{1},j_{2},\dots ,j_{N_{a}}\}$ is a permutation of $\{1,2,\dots
,N_{a}\}$ such that
\begin{equation}
\pi ^{-1}(j_{1})<\pi ^{-1}(j_{2})<\dots <\pi ^{-1}(j_{N_{a}}),  \label{r37}
\end{equation}
and $\{l_{1},l_{2},\dots ,l_{N_{b}}\}$ is a permutation of $\{1,2,\dots
,N_{b}\}$ such that
\begin{equation}
\pi ^{-1}(N_{a}+l_{1})<\pi ^{-1}(N_{a}+l_{2})<\dots <\pi
^{-1}(N_{a}+l_{N_{b}}),  \label{r38}
\end{equation}
is, according to the definition in \cite{G2002}, totally non negative, i.e. has all its minors non negative.
\end{theorem}
\textsl{Proof.\/} From (\ref{r33}) and (\ref{r34}) we have
\begin{equation}
\Do_{ml}=\mathcal{B}_{\pi (m),l}  \label{r12}
\end{equation}
Let, now, $J$ be the set
\begin{equation}
J=\{\pi ^{-1}(N_{a}+1),\dots ,\pi ^{-1}(\mathcal{N})\}  \label{r13}
\end{equation}
and let
\begin{equation}
\mathcal{C}J=\{\pi ^{-1}(1),\dots ,\pi ^{-1}(N_{a})  \label{r14}
\end{equation}
be its complement in the set $\{1,\ldots ,\No\}$.

Then, from (\ref{r7}) we have
\begin{align}
& \sum_{l=1}^{N_{b}}\chi \left( i\kappa _{\pi ^{-1}(N_{a}+l)}\right) \left[
e^{K_{\pi ^{-1}(N_{a}+l)}}\delta _{l,l^{\prime }}+i\sum_{j=1}^{N_{a}}\dfrac{
\Delta \left( i\kappa _{\pi ^{-1}(j)}\right) e^{K_{\pi ^{-1}(j)}}}{\left(
\kappa _{\pi ^{-1}(N_{a}+l)}-\kappa _{\pi ^{-1}(j)}\right) \Delta ^{\prime
}\left( i\kappa _{\pi ^{-1}(N_{a}+l)}\right) }d_{j,l^{\prime }}\right] =
\notag \\
& \qquad =-\sum_{j=1}^{N_{a}}\Delta \left( i\kappa _{\pi ^{-1}(j)}\right)
e^{K_{\kappa _{\pi ^{-1}(j)}}}d_{j,l^{\prime }}.  \label{r16}
\end{align}
For the $\tau $ function we obtain, therefore,
\begin{equation}
\tau (x)=\det \left( \delta _{l,l^{\prime }}+i\sum_{j=1}^{N_{a}}\dfrac{
\Delta \left( i\kappa _{\pi ^{-1}(j)}\right) e^{K_{\pi ^{-1}(j)}-K_{\pi
^{-1}(N_{a}+l)}}}{\left( \kappa _{\pi ^{-1}(N_{a}+l)}-\kappa _{\pi
^{-1}(j)}\right) \Delta ^{\prime }\left( i\kappa _{\pi
^{-1}(N_{a}+l)}\right) }d_{j,l^{\prime }}\right) _{l,l^{\prime }=1,\ldots
,N_{b}}.  \label{r17}
\end{equation}
Let us re-order the rows in the matrix in the determinant from $l=1,2,\dots
,N_{b}$ to a permutation $l_{1},l_{2},\dots ,l_{N_{b}}$ such that
\begin{equation}
\kappa _{\pi ^{-1}(N_{a}+l_{1})}<\kappa _{\pi ^{-1}(N_{a}+l_{2})}<\dots
<\kappa _{\pi ^{-1}(N_{a}+l_{N_{b}})}  \label{r18}
\end{equation}
and let us re-order the sum over $j$ from $j=1,2,\dots ,N_{a}$ to a
permutation $j_{1},j_{2},\dots ,j_{N_{a}}$ such that
\begin{equation}
\kappa _{\pi ^{-1}(j_{1})}<\kappa _{\pi ^{-1}(j_{2})}<\dots <\kappa _{\pi
^{-1}(j_{N_{a}})}.  \label{r19}
\end{equation}
Then, up to an unessential sign, the $\tau$ function can be rewritten as
\begin{align}
\tau (x)&=\det\Biggl( E_{r,l^{\prime }}+\nonumber\\
&+\sum_{k=1}^{N_{a}}\left( \frac{
ie^{-K_{\pi ^{-1}(N_{a}+l_{r})}}}{\Delta ^{\prime }\left( i\kappa _{\pi
^{-1}(N_{a}+l_{r})}\right) }\right) \Lambda _{rk}\left( \Delta \left(
i\kappa _{\pi ^{-1}(j_{k})}\right) e^{K_{\pi ^{-1}(j_{k})}}\right)
d_{j_{k},l^{\prime }}\Biggr) _{r,l^{\prime }=1,\ldots ,N_{b}},  \label{r20}
\end{align}
where
\begin{align}
E_{r,l^{\prime }}& =\delta _{l_{r},l^{\prime }}  \label{r21} \\
\Lambda _{rk}& =\frac{1}{\kappa _{\pi ^{-1}(N_{a}+l_{r})}-\kappa _{\pi
^{-1}(j_{k})}}.  \label{r22}
\end{align}
If we multiply from the right the matrix in the determinant by $E^{-1}$ we
get up to a sign
\begin{equation}
\tau (x)=\det (E_{N_{b}}+T)  \label{r23}
\end{equation}
where ($r,s=1,\dots ,N_{b}$)
\begin{equation}
T_{rs}=\sum_{k=1}^{N_{a}}\left( \frac{ie^{-K_{\pi ^{-1}(N_{a}+l_{r})}}}{
\Delta ^{\prime }\left( i\kappa _{\pi ^{-1}(N_{a}+l_{r})}\right) }\right)
\Lambda _{rk}\left( \Delta \left( i\kappa _{\pi ^{-1}(j_{k})}\right)
e^{K_{\pi ^{-1}(j_{k})}}\right) d_{j_{k},l_{s}}.  \label{r24}
\end{equation}
By using repeatedly the Binet-Cauchy formula, we have
\begin{equation}
\tau (x)=\sum_{n=0}^{N_{b}}\sum_{1\leq r_{1}<r_{2}<\dots <r_{n}\leq
N_{b}}T\left(
\begin{array}{l}
r_{1},r_{2},\dots ,r_{n} \\
r_{1},r_{2},\dots ,r_{n}
\end{array}
\right) .  \label{r25}
\end{equation}
and, then,
\begin{align}
& T\left(
\begin{array}{l}
r_{1},r_{2},\dots ,r_{n} \\
r_{1},r_{2},\dots ,r_{n}
\end{array}
\right) =\sum_{1\leq k_{1}<k_{2}<\dots <k_{n}\leq N_{a}}\left(
\prod_{m=1}^{n}\frac{ie^{-K_{\pi ^{-1}(N_{a}+l_{r_{m}})}}}{\Delta ^{\prime
}\left( i\kappa _{\pi ^{-1}(N_{a}+l_{r_{m}})}\right) }\right) \times  \notag
\\
& \quad \times \left( \prod_{m=1}^{n}\Delta \left( i\kappa _{\pi
^{-1}(j_{k_{m}})}\right) e^{K_{\pi ^{-1}(j_{k_{m}})}}\right) \Lambda \left(
\begin{array}{l}
r_{1},r_{2},\dots ,r_{n} \\
k_{1},k_{2},\dots ,k_{n}
\end{array}
\right) d\left(
\begin{array}{l}
k_{1},k_{2},\dots ,k_{n} \\
r_{1},r_{2},\dots ,r_{n}
\end{array}
\right) .  \label{r26}
\end{align}
Now, if we impose that the coefficients of the exponents are non negative, we get
\begin{align}
& \det E\left( \prod_{m=1}^{n}\frac{i}{\Delta ^{\prime }\left( i\kappa
_{\pi ^{-1}(N_{a}+l_{r_{m}})}\right) }\right) \left( \prod_{m=1}^{n}\Delta
\left( i\kappa _{\pi ^{-1}(j_{k_{m}})}\right) \right) \times  \notag \\
& \qquad \times \Lambda \left(
\begin{array}{l}
r_{1},r_{2},\dots ,r_{n} \\
k_{1},k_{2},\dots ,k_{n}
\end{array}
\right) d\left(
\begin{array}{l}
k_{1},k_{2},\dots ,k_{n} \\
r_{1},r_{2},\dots ,r_{n}
\end{array}
\right) \geq 0.  \label{r27}
\end{align}

Since
\begin{equation}
\Lambda \left(
\begin{array}{l}
r_{1},r_{2},\dots ,r_{n} \\
k_{1},k_{2},\dots ,k_{n}
\end{array}
\right) =\dfrac{\displaystyle\prod_{1\leq i<m\leq n}
\left(\kappa_{\pi^{-1}(N_{a}+l_{r_{i}})}-\kappa_{\pi^{-1}(N_{a}+l_{r_{m}})}\right)
\left(\kappa_{\pi ^{-1}(j_{k_{m}})}-\kappa _{\pi ^{-1}(j_{k_{i}})}\right) }{\displaystyle\prod_{i,m=1}^{n}
\left(\kappa _{\pi ^{-1}(N_{a}+l_{r_{i}})}-\kappa _{\pi^{-1}(j_{k_{m}})}\right)}  \label{r28}
\end{equation}
the denominator in (\ref{r28}) cancel with $\prod_{m=1}^{n}\Delta(i\kappa _{\pi ^{-1}(j_{k_{m}})})$ in (\ref{r27}), and, after reversing the order of the $r_{i}$'s, we have
\begin{equation}
\left(\prod_{m=1}^{n}\Delta(i\kappa_{\pi ^{-1}(j_{k_{m}})})\right)
\Lambda\left(\begin{array}{l}
r_{n},r_{2},\dots ,r_{1} \\
k_{1},k_{2},\dots ,k_{n}
\end{array}\right) >0.  \label{r29}
\end{equation}
Finally, since
\begin{equation}
-i\Delta^{\prime }(i\kappa _{\pi ^{-1}(N_{a}+l_{r_{m}})})
=(-1)^{l_{r_{m}}-1}\displaystyle\prod_{s=1,s\neq m}^{n}\left\vert -\kappa _{\pi^{-1}(N_{a}+l_{r_{m}})}+
\kappa _{\pi ^{-1}(N_{a}+l_{r_{s}})}\right\vert,\label{r30}
\end{equation}
we get the condition
\begin{equation}
(-1)^{\sum_{m=1}^{n}(1+l_{m})}d\left(
\begin{array}{l}
k_{n},k_{n-1},\dots ,k_{1} \\
r_{1},r_{2},\dots ,r_{n}
\end{array}
\right) \geq 0.  \label{r31}
\end{equation}
We conclude that the matrix
\begin{equation}
\widetilde{d}_{kr}=(-1)^{1+l_{r}}d_{\pi (j_{(N_{a}+1-k)}),l_{r}},\qquad
k=1,\dots ,N_{a},\quad r=1,\dots ,N_{b}  \label{r32}
\end{equation}
is totally non negative. $\blacksquare$

From this theorem the following Corollary follows easily.
\begin{corollary}
A $N_{a}\times N_{b}$ matrix $\widetilde{d}$ is totally non negative, i.e.
has all its minors non negative, if and only if the $\No\times
N_{b}$ matrix
\begin{equation}
\mathcal{D}=\left(
\begin{array}{l}
d \\
E_{b}
\end{array}
\right) ,  \label{r39}
\end{equation}
where
\begin{equation}
d_{jl}=(-1)^{l+1}\widetilde{d}_{N_{a}+1-j,l},\qquad j=1,\dots ,N_{a},\quad
l=1,\dots ,N_{b},  \label{r40}
\end{equation}
and $E_{b}$ is the unit $N_{b}\times N_{b}$ matrix, has all its maximal
minors non negative.$\blacksquare $
\end{corollary}

\end{document}